\documentclass[11pt,reqno,final]{amsart}\usepackage[]{graphicx}\usepackage[]{color}
\makeatletter
\def\maxwidth{ %
  \ifdim\Gin@nat@width>\linewidth
    \linewidth
  \else
    \Gin@nat@width
  \fi
}
\makeatother

\definecolor{fgcolor}{rgb}{0.345, 0.345, 0.345}

\usepackage{framed}
\makeatletter
 {\par\unskip\endMakeFramed%
 \at@end@of@kframe}
\makeatother

\definecolor{shadecolor}{rgb}{.97, .97, .97}
\definecolor{messagecolor}{rgb}{0, 0, 0}
\definecolor{warningcolor}{rgb}{1, 0, 1}
\definecolor{errorcolor}{rgb}{1, 0, 0}
\newenvironment{knitrout}{}{} 

\usepackage{alltt}
\usepackage{graphicx}
\usepackage{color}
\usepackage{times}
\usepackage[neverdecrease]{paralist}
\usepackage[round,elide]{natbib}
\usepackage{amssymb}
\usepackage{mathrsfs}
\usepackage{eucal}
\usepackage{mathtools}
\usepackage[colorlinks=true]{hyperref}

\numberwithin{equation}{section}
\renewcommand\thesection{\arabic{section}}
\renewcommand\thesubsection{\thesection.\arabic{subsection}}

\newcommand\eps{\varepsilon}
\newcommand\prob[1]{\mathbb{P}\left[{#1}\right]}
\newcommand\expect[1]{\mathbb{E}\left[{#1}\right]}
\newcommand\indicator[1]{\mathbb{I}_{{#1}}}

\newcommand{\dd}[1]{\mathrm{d}{#1}}

\newcommand{\pd}[3][]{%
  \def\ord{#1} \ifx\ord\empty%
  \frac{\partial{#2}}{\partial{#3}}%
  \else \frac{\partial^{#1}{#2}}{\partial{#3}^{#1}}%
  \fi
}
\newcommand{\deriv}[3][]{%
  \def\ord{#1} \ifx\ord\empty%
  \frac{\dd{#2}}{\dd{#3}}%
  \else \frac{\dd^{#1}{#2}}{\dd{#3}^{#1}}%
  \fi
}
\newcommand{\opnorm}[1]{{\left\vert\kern-0.25ex\left\vert\kern-0.25ex\left\vert{#1}
        \right\vert\kern-0.25ex\right\vert\kern-0.25ex\right\vert}}

\newcommand\sv[1]{\mathbf{#1}}
\newcommand\sg[1]{\mathcal{#1}}
\newcommand\Sp[1]{\mathbb{#1}}

\usepackage[small,compact]{titlesec}
\newcommand\periodafter[1]{{#1}.}
\titleformat{\section}[hang]{\large\bfseries}{\periodafter\thesection}{2ex}{\periodafter}{}
\titleformat{\subsection}[hang]{\normalsize\bfseries}{\periodafter\thesubsection}{1ex}{\periodafter}{}
\titleformat{\subsubsection}[runin]{\normalsize\bfseries}{\thesubsubsection}{0em}{\periodafter}{}
\titleformat{\paragraph}[runin]{\normalsize\bfseries}{\theparagraph}{0em}{\periodafter}{}
\titlespacing{\section}{0em}{*1}{*0}
\titlespacing{\subsection}{0em}{*0}{*0}
\titlespacing{\paragraph}{0em}{*1}{*1}

\usepackage[sort&compress]{cleveref}

\theoremstyle{plain}
\newtheorem{thm}{Theorem}
\newtheorem{corol}[thm]{Corollary}
\newtheorem{prop}[thm]{Proposition}
\newtheorem{lemma}[thm]{Lemma}

\theoremstyle{definition}

\theoremstyle{remark}

\crefname{figure}{Fig.}{Figs.}
\Crefname{figure}{Fig.}{Figs.}
\crefname{table}{Table}{Tables}
\Crefname{table}{Table}{Tables}
\crefname{equation}{Eq.}{Eqs.}
\Crefname{equation}{Eq.}{Eqs.}
\creflabelformat{equation}{#2#1#3}
\crefname{appendix}{Appendix}{Appendices}
\Crefname{appendix}{Appendix}{Appendices}
\crefname{chapter}{Ch.}{Chs.}
\Crefname{chapter}{Ch.}{Chs.}
\crefname{section}{\S}{\S\S}
\Crefname{section}{\S}{\S\S}
\crefformat{section}{#2\S#1#3}
\Crefformat{section}{#2\S#1#3}
\crefmultiformat{section}{#2\S\S#1#3}{ and~#2#1#3}{, #2#1#3}{, and~#2#1#3}
\Crefmultiformat{section}{#2\S\S#1#3}{ and~#2#1#3}{, #2#1#3}{, and~#2#1#3}
\crefname{thm}{Theorem}{Theorems}
\Crefname{thm}{Theorem}{Theorems}
\crefname{corol}{Corollary}{Corollaries}
\Crefname{corol}{Corollary}{Corollaries}
\crefname{prop}{Proposition}{Propositions}
\Crefname{prop}{Proposition}{Propositions}
\crefname{conj}{Conjecture}{Conjectures}
\Crefname{conj}{Conjecture}{Conjectures}
\crefname{lemma}{Lemma}{Lemmas}
\Crefname{lemma}{Lemma}{Lemmas}
\crefname{defn}{Definition}{Definitions}
\Crefname{defn}{Definition}{Definitions}

\usepackage{algorithm}
\usepackage{algorithmic}

\definecolor{darkgreen}{rgb}{0.0,0.5,0.2}

\definecolor{darkpurple}{rgb}{0.5,0.0,1.0}

\definecolor{darkbrown}{rgb}{0.8,0.8,0.0}

\title[Sampled Moran genealogy process]{The Sampled Moran Genealogy Process}
\author[King]{Aaron~A.~King}
\address{
  A.~A.~King,
  Department of Ecology \& Evolutionary Biology,
  Center for the Study of Complex Systems,
  Center for Computational Medicine \& Biology, and
  Michigan Institute for Data Science,
  University of Michigan,
  Ann Arbor, MI 48109 USA
}
\email{kingaa@umich.edu}
\urladdr{\href{https://kinglab.eeb.lsa.umich.edu/}{https://kinglab.eeb.lsa.umich.edu/}}
\author[Lin]{Qianying Lin}
\address{
  Q.-Y. Lin,
  Michigan Institute for Data Science,
  University of Michigan,
  Ann Arbor, MI 48109 USA
}
\author[Ionides]{Edward~L.~Ionides}
\address{
  E.~L.~Ionides,
  Department of Statistics and
  Michigan Institute for Data Science,
  University of Michigan,
  Ann Arbor, MI 48109 USA
}
\date{\today}

\hypersetup{pdftitle={The Moran Genealogy Process}}
\hypersetup{pdfauthor={A.A. King, Q.-Y. Lin, E.L. Ionides}}
\hypersetup{urlcolor=blue,citecolor=blue,linkcolor=blue,filecolor=blue}

\IfFileExists{upquote.sty}{\usepackage{upquote}}{}
\begin{document}

\begin{abstract}
  We define the Sampled Moran Genealogy Process, a continuous-time Markov process on the space of genealogies with the demography of the classical Moran process, sampled through time.
  To do so, we begin by defining the Moran Genealogy Process using a novel representation.
  We then extend this process to include sampling through time.
  We derive exact conditional and marginal probability distributions for the sampled process under a stationarity assumption, and an exact expression for the likelihood of any sequence of genealogies it generates.
  This leads to some interesting observations pertinent to existing phylodynamic methods in the literature.
  Throughout, our proofs are original and make use of strictly forward-in-time calculations and are exact for all population sizes and sampling processes.
\end{abstract}

\maketitle

\section{Introduction}

The Moran process \citep{Moran1958} plays an important role in the theory of population genetics and is intimately related to Kingman's \citeyearpar{Kingman1982,Kingman1982c,Kingman1982b} coalescent process, itself a foundational component of modern population genetics, phylogenetics, and phylodynamics \citep{Hudson1990,Donnelly1995,Stephens2000,Rosenberg2002,Ewens2004,Volz2009a,Rasmussen2011}.
Kingman formulated the coalescent as a backward-in-time Markov process whereby genealogical lineages randomly coalesce with one another.
He made explicit connections with the classical Moran model, and connections exist with a broader collection of population genetics models \citep{Cannings1974,Ewens2004,Moehle2000,Moehle2010,Etheridge2011,Etheridge2019}.

Much of the literature on coalescent theory is focused on changes in the frequencies of alleles under various assumptions regarding population dynamics, natural selection, and genetic drift \citep{Hein2005,Durrett2008a,Wakeley2008}.
Genealogies and the coalescent play a different role in the phylodynamics literature.
Broadly, phylodynamics is the attempt to infer determinants of pathogen transmission and evolution on the basis of pathogen genome sequences collected through time \citep{Grenfell2004,Frost2015,Smith2017}.
Currently, major phylodynamic approaches view genealogies derived from such sequences as static representations of the history of transmission events;
deterministic or stochastic transmission models are then fit to genealogies using backward-in-time arguments that rely on various approximations, including large population size, small sample fraction, and, in some cases, time-reversibility of the transmission process.
Examples of these approaches can be found in the papers of \citet{Gernhard2008}, \citet{Volz2009a,Volz2013,Volz2014}, \citet{Rasmussen2011,Rasmussen2014a}, \citet{Stadler2013}, and \citet{duPlessis2015}.

Motivated by the desire to make phylodynamic arguments more rigorous and to dispense with unnecessary approximations, we refocus the discussion onto the evolution of the genealogy that, at any given time, describes the full set of relationships among the members of a population alive at that time.
In particular, we view the genealogy as a dynamic object and seek to describe the stochastic processes that generate it.
Accordingly, in \cref{sec:MGP}, we define the Moran Genealogy Process (MGP), a continuous-time Markov process that take values in the space of genealogies with real-valued branch lengths.
To aid in visualization and reasoning about this process, we introduce a novel representation of the MGP, as a parlor game for $n$ players.
We state the most important properties about this process---its uniform ergodicity, the form of its invariant distribution, and its projective symmetry---postponing the proofs to an appendix.
Although these properties are exact analogues of well known properties of the Moran process and the Kingman coalescent, the proofs are novel, constructive, and strictly forward-in-time.
In \cref{sec:SMGP}, we extend the MGP to include asynchronous sampling.
We give a parlor-game representation to complement the formal definition, and go on to derive exact expressions for the conditional and marginal distributions the observable genealogies generated by this process.
Finally, in \cref{sec:discussion}, we indicate some implications for some existing phylodynamic approaches.
In particular, the nature of the approximations made in these approaches is clarified at the same time that the need for them is diminished.
An examination of the proofs reveals that these results will generalize to a broad class of birth-death processes.

It appears that \citet{Pfaffelhuber2006}, \citet{Pfaffelhuber2011}, and \citet{Greven2013} were the first to conceive of genealogy-valued stochastic processes and to derive many of their important properties.
In particular, \citet{Pfaffelhuber2006} and \citet{Pfaffelhuber2011} established the nature of the tree-depth and tree-length processes, respectively, in the limit of large population size.
\citet{Greven2013}, represented the evolving genealogy as a Markov process on a topological space, the points of which are themselves metric spaces, and proved the well-posedness of the martingale problem for tree-valued resampling dynamics under both the finite-population Moran process and its infinite population-size limit, a Fleming-Viot diffusion.
Their results presage those of \cref{sec:MGP} below, which are formulated, however, in terms of our concrete parlor-game representation, and which require no large population-size approximations.
The work of \citet{Wirtz2019} offers another recent perspective on some of these issues, though these authors confined themselves to the discrete aspects of genealogies, whereas in this paper, we consider both the discrete and continuous (i.e., branch-length) aspects of genealogies.
The concept of an evolving genealogy also figured in the work of \citet{Smith2017}, who used it as a basis for a computational approach to phylodynamic inference.

\section{The Moran Genealogy Process}\label{sec:MGP}

\subsection{A Markov process in the space of genealogies}

\begin{figure}[h]
  \centering
\begin{knitrout}\small
\definecolor{shadecolor}{rgb}{0.969, 0.969, 0.969}\color{fgcolor}

{\centering \includegraphics[width=0.95\linewidth]{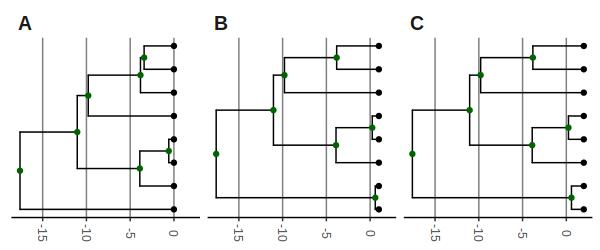} 

}

\end{knitrout}
  \caption{
    \textbf{The Moran Genealogy Process.}
    Three equally-spaced instants in a realization of the Moran genealogy process (MGP) of size $n=8$.
    The MGP is a continuous-time process on tip-labelled genealogies with branch lengths.
    In the MGP, tips of the genealogy (corresponding to living members of the population) face a constant event hazard.
    At each event, a randomly selected individual gives birth and a second random individual dies.
    Accordingly, the genealogy grows continuously at its leading (right) edge, while internal nodes drop in discrete events as ancestors are ``forgotten''.
    Between panels \textbf{A} and \textbf{B}, two Moran (birth+death) events have occurred;
    between panels \textbf{B} and \textbf{C}, zero events have occurred.
    \label{fig:gtree}
  }
\end{figure}

The name of P.~A.~P.~Moran has been associated with a number of related stochastic processes arising in population genetics.
These processes share the common features that they involve a finite population of identical, asexual individuals who reproduce and die stochastically in continuous time.
The population size is kept deterministically constant by requiring that each reproduction event is coincident with a death event.
Such models are closely related to the coalescent of \citet{Kingman1982,Kingman1982c,Kingman1982b}, which plays a prominent role in population genetics and phylogenetics.
We explore these connections from a new angle by defining the \emph{Moran Genealogy Process} (MGP), a stochastic process on the space of genealogies.
In common with other models bearing Moran's name, we make the assumptions that
\begin{inparaenum}[(a)]
\item the process is a continuous-time Markov process, with a constant event rate, and
\item at each event, one asexual individual gives birth and another dies, so that the population size, $n$, remains constant.
\end{inparaenum}
At any particular time, the state of the MGP is a genealogy---a tree with branch lengths---that relates the $n$ living members of the population via their shared ancestry.
This consists of links between living individuals and those past individuals who are the most recent common ancestors of sets of currently living individuals.
As living individuals reproduce, the genealogy grows at its leading edge;
it dissolves at its trailing edge, as ancestors are ``forgotten''.
\Cref{fig:gtree} illustrates;
see also the animations in the \href{https://kinglab.eeb.lsa.umich.edu/kingaa/mgp/moran.html#one_simulation_of_the_mgp}{online appendix}.

The genealogies of the MGP have two aspects, one discrete, the other continuous.
The discrete aspect, encoding the topological evolution of the genealogies, evolves as a jump process:
at each event, a new branch appears at the leading edge and an internal node is dropped.
The continuous aspect, tracking the dynamics of branch lengths, evolves continuously between event times as the latter grow linearly at the leading edge of the tree and discontinuously at event times as internal nodes are dropped.

In this paper, we use the term \emph{chain} to refer to any discrete-time stochastic process, regardless of the nature of its state space.

\subsection{The Moran genealogy game}

Although genealogies are naturally represented using trees, these representations are not unique, and it can be challenging to reason about their properties.
Accordingly, we map the MGP onto a parlor game for which intuition is more readily available.
The game players will represent the internal nodes of the genealogy.
Each player will hold two balls representing the two child nodes.
The players will be seated in a row of chairs, from left (past) to right (future).
Each configuration of seated players will correspond uniquely to a state of the MGP.
We will give a formal definition, using the game terminology, after we describe the game itself.
The Moran genealogy process in a population of size $n$, then, is equivalent to the following parlor game for $n$ players.

\paragraph{Equipment}

We have $n$ black balls, numbered $1\,\dots,n$, and $n$ green balls, each one of which is inscribed with the name of one of the players (we assume the names are unique).
Each player receives a slate, the green ball bearing his or her name, and a randomly chosen black ball.
We arrange $n$ seats in a row and number them $0$ through $n-1$.
We also have a clock which, when started, runs for a random amount of time and then stops.
The stopping times are exponentially distributed with rate parameter $\mu$.

\paragraph{Setup}

To begin the game, an arbitrarily chosen player takes seat number 0.
Upon her slate, she writes ``$-\infty$''.
The remaining players are then seated sequentially, from left to right, in arbitrary order.
As each successive player takes a seat, he exchanges his green ball for a randomly selected black ball held by one of the already-seated players.
He then writes a real number upon the slate;
the only constraint on his choice is that it be at least as large as the number on the slate of the player to her left yet not greater than zero.
Thus, the player taking seat $m$ encounters the following situation.
The $m$ players already seated hold among themselves $m$ green balls and $m$ black balls.
The player to be seated therefore has $m$ choices as to the black ball he will exchange for his green ball.
The leftmost player holds two green balls and the rightmost, two black balls;
the other players may have one of each color.
Each player's green ball (other than that of the leftmost player) is held by a player seated to her left.

\paragraph{Play}

\begin{figure}
\begin{knitrout}\small
\definecolor{shadecolor}{rgb}{0.969, 0.969, 0.969}\color{fgcolor}

{\centering \includegraphics[width=0.95\linewidth]{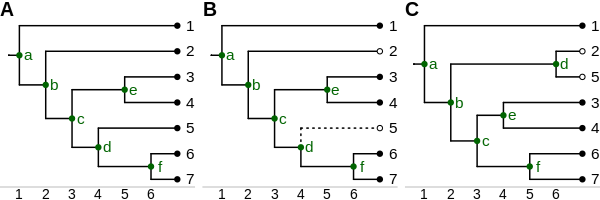} 

}

\end{knitrout}
  \caption[The $\sv{W}$ chain.]{
    \textbf{The $\sv{W}$ chain.}
    Here, $n=7$.
    \textbf{(A)}~The $n-1=6$ players, seated in seats 1--6 and named \textbf{a}, \dots, \textbf{f} (green labels), represent the internal nodes.
    Each player holds two balls, each of which may be green or black.
    The green balls are in 1-1 correspondence with players and each has the name of a player inscribed upon it.
    A player holding a green ball is the immediate ancestor of the named player.
    The $n$ black balls are numbered and represent members of the extant population and leaves of the genealogy.
    Thus player \textbf{b} holds black ball number 2 and green ball c, while \textbf{e} holds the black balls 3 and 4.
    The player in seat 0 represents the distant ancestor at time $-\infty$ and can be visualized as a single lineage extending infinitely far to the left from the root.
    Since this player never moves and conveys only redundant information, he is omitted from the diagrams and from the state space.
    \textbf{(B)}~In each round of play, an ordered pair of black balls is selected at random.
    In the illustrated case, the first is ball 5, held by player \textbf{d}.
    Accordingly, player \textbf{d} exchanges green ball f with player \textbf{c} for green ball d and moves to the rightmost position (position 6).
    Players \textbf{e} and \textbf{f} each shift one position to the left.
    The second ball selected is ball 2, held by player \textbf{b}, who exchanges ball 2 with player \textbf{d} for ball d.
    The resulting configuration is shown in panel~\textbf{C}.
    \label{fig:wchain}
  }
\end{figure}

Play proceeds in rounds (\cref{fig:wchain}).
Each round begins when the clock is started.
When it stops, two black balls are chosen at random (without replacement).
The player---call him X---holding the first black ball, stands up.
Player X exchanges his second ball for the green ball bearing his name.
This will always be held by a player seated to his left.
All players to the right of X then shift one seat to the left, leaving the rightmost seat empty.
Player X is said to have been ``killed''.
Next, the player, Y, holding the second randomly-selected black ball, trades it for X's green ball. 
Player X now takes the rightmost seat and writes the current time upon his slate.
Player Y is said to have ``given birth'' and player X to have been ``reborn''.
We sometimes refer to the conjoined birth and death events as a single ``Moran event''.

Note that, since the player in seat $0$ always holds two green balls (one of them her own) and never a black ball, she can never be killed and therefore remains in this seat throughout the game.
Nor does the number on her slate ever change.

\paragraph{Relation to the Moran genealogy process (MGP)}

The correspondence between the game and genealogies is as follows.
The black balls correspond to individuals in the extant population, i.e., to the tips of the genealogical tree.
The seats numbered $1,\dots,n-1$ (that is, all but the leftmost seat) correspond to the time-ordered $n-1$ internal nodes, seat number $1$ being the root, i.e., the most recent common ancestor of all extant individuals.
The green balls record the topology of the tree:
each player holding a green ball is the immediate parent of the player named on that ball.
\Cref{fig:gtree} illustrates.

Note that, at every stage of the game, the arrangement of seated players has \textbf{Property G}:
\begin{compactenum}[(i)]
\item ordered from left to right, the times on the slates are non-decreasing,
\item each player holds exactly two balls,
\item no green ball is held by a player seated to the right of the player named on the ball, and
\item the only player holding her own ball is the player in seat 0.
\end{compactenum}
\label{propertyG}
It is straightforward to show that every arrangement of seated players having Property G corresponds to a genealogy, and vice versa.

\paragraph{Formal definition}

The MGP is defined to be a continuous-time Markov process, with c\`adl\`ag sample paths, on the space of tip-labeled genealogies with branch lengths and unordered descendants.
The topological structure of a tree at time $t$ is represented by the list $\sv{W}(t) = (\sv{W}_1(t), \dots, \sv{W}_{n-1}(t))$, where each $\sv{W}_m(t)$ is the (unordered) pair of balls held by the player in seat $m$.
Note that $\sv{W}_0$ is left out of $\sv{W}$:
the balls held by player $0$---who represents the distant ancestor---convey only redundant information.
Let $\Sp{W}^n$ be the finite set of all such states.
In \cref{sec:w-size}, it is shown that $\vert\Sp{W}^n\vert=\prod_{m=1}^{n-1}\!\binom{m+1}{2}=n!\,(n-1)!/2^{n-1}$.

The number written on a player's slate is the time at which that player was most recently seated.
Let $\sv{T}(t)=(T_1(t),\dots,T_{n-1}(t))$ be the vector of numbers on the slates at time $t$, ordered left to right.
Let $\sv{S}(t)=(\sv{S}_1(t),\dots,\sv{S}_{n-1}(t))$, where for $m<n-1$, $\sv{S}_m(t)=T_{m+1}(t)-T_{m}(t)$ and $\sv{S}_{n-1}(t)=t-T_{n-1}(t)$.
Then the $\sv{S}_m(t)$ are the durations of the \emph{coalescent intervals}, i.e., the intervals between successive branch points when the latter are ordered in time.
The state space of the MGP is defined to be $\Sp{X}^n=\Sp{W}^n\times\Sp{S}^n$, where $\Sp{S}^n\coloneqq\Sp{R}_{+}^{n-1}$ and the MGP itself can be written $\sv{X}(t)=(\sv{W}(t),\sv{S}(t))$, for $t\ge 0$.
\Cref{fig:gtree} depicts the MGP;
animated illustrations are available in the \href{https://kinglab.eeb.lsa.umich.edu/kingaa/mgp/moran.html#one_simulation_of_the_mgp}{online appendix}.
\Cref{fig:wchain} illustrates the projection of the embedded chain onto $\Sp{W}^n$.

\subsection{Dynamics of the MGP}

In \cref{sec:properties}, we derive the main properties of the MGP from first principles.
The main result is contained in the following, which summarizes \cref{thm:ergodic,thm:pi}.

\begin{thm}
  \label{thm:main_mgp}
  The Moran genealogy process is uniformly and exponentially ergodic with the unique invariant measure
  \begin{equation*}
    \pi_n(\dd{\sv{w}}\,\dd{\sv{s}}) = \left({{\mu}\over{\smash{\binom{n}{2}}}}\right)^{n-1}\,\exp\left(-\sum_{j=1}^{n-1}\!\frac{\binom{j+1}{2}}{\binom{n}{2}}\,\mu\,\sv{s}_j\right)\,\dd{\sv{w}}\,\dd{\sv{s}},
  \end{equation*}
  where $\dd{\sv{w}}$ is the counting measure on $\Sp{W}^n$ and $\dd{\sv{s}}$ is Lebesgue measure on $\Sp{S}^n$.
\end{thm}

Thus, from any initial configuration, the state of the MGP converges to the invariant measure indicated, which is the probability measure of the Kingman $n$-coalescent \citep{Kingman1982,Kingman1982c,Kingman1982b}.
The form of this invariant measure is well known from coalescent theory, since the Kingman $n$-coalescent is dual to the Moran process.
This can be easily established from backward-in-time arguments using the exchangeability of the offspring distributions of the $n$ members of the population.
However, the proofs in \cref{sec:properties} are of interest in that they prove ergodicity and explicitly calculate the form of the invariant measure using strictly forward-looking arguments.

The convergence of \cref{thm:main_mgp} should not be confused with convergence in the limit of large population size ($n\to\infty$).
For example, \citet{Moehle2000} showed, for a broad class of processes including the Moran process, that the invariant probability distribution approaches that of the Kingman coalescent as the population size tends to infinity \citep[see also][]{Berestycki2009}.
By contrast, \cref{thm:main_mgp} shows that this limit holds for every finite $n$ as $t\to\infty$, uniformly in the initial conditions.

\subsection{Synchronous sampling}\label{sec:syncsamp}

\begin{figure}
\begin{knitrout}\small
\definecolor{shadecolor}{rgb}{0.969, 0.969, 0.969}\color{fgcolor}

{\centering \includegraphics[width=0.95\linewidth]{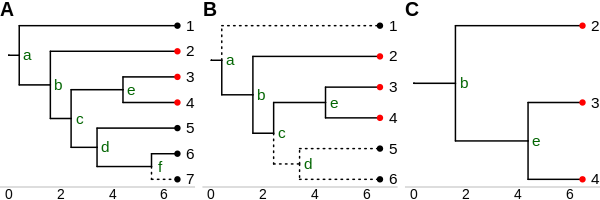} 

}

\end{knitrout}
  \caption{
    \textbf{The pruning procedure for synchronous sampling.}
    Here, $n=7$ and the six players seated in seats 1--6 are named \textbf{a}--\textbf{f}, as in \cref{fig:wchain}.
    The sample size $k=3$.
    The horizontal axis shows time.
    In panel \textbf{A}, $k$ black balls have been selected at random and replaced with red balls (numbers 2, 3, and 4).
    The highest-numbered black ball (number 7) is to be removed.
    Player \textbf{f} holds black ball 7, so she gives her other ball (black ball 6) to player \textbf{d} in exchange for the ball bearing her name.
    She then departs (\textbf{B}).
    The same procedures are repeated for the remaining black balls (numbers 1, 5, and 6), with the result that players \textbf{d}, \textbf{c}, and \textbf{a} are dismissed, in that order.
    The resulting genealogy (\textbf{C}) spans only the sampled individuals.
    \label{fig:sync_prune}
  }
\end{figure}

Suppose one randomly samples $k$ of the $n$ individuals in the population: what is the genealogy relating them?
We can represent the process of sampling a subgenealogy of size $k$ in terms of the Moran genealogy game as follows.
We randomly select $k$ black balls and replace them with red balls.
Then we perform the following iterative \emph{pruning} procedure (\cref{fig:sync_prune}):
\begin{inparaenum}[(1)]
\item The player holding the highest-numbered black ball stands up.
\item He exchanges his second ball for the green ball bearing his name.
\item The standing player is dismissed, all players seated to his right shift one seat to the left, and the rightmost chair is removed.
\end{inparaenum}
At each iteration of this procedure, we remove the highest-numbered black ball from play and dismiss one player, preserving Property G (\cpageref{propertyG}) all the while.
Thus after $n-k$ steps, the configuration of the balls held by the remaining $k$ players, and the numbers written on their slates, together determine the sampled genealogy (\cref{fig:sync_prune}C).
The proof of the following theorem can be found in \cref{sec:sampling}.

\begin{thm}\label{thm:synsamp}
  Let $\sv{X}_n$ be the stationary Moran genealogy process of size $n$ and event rate $\mu$.
  For $k\le n$, let $\sv{Z}_k$ be the corresponding size-$k$ sampled process.
  Then the marginal probability distribution of $\sv{Z}_k(t)$ on $\Sp{X}^k$ is given by the measure
  \begin{equation*}
    \left({{\mu}\over{\smash{\binom{n}{2}}}}\right)^{k-1}\,\exp\left(-\sum_{j=1}^{k-1}\!\frac{\binom{j+1}{2}}{\binom{n}{2}}\,\mu\,\sv{s}_j\right)\,\dd{\sv{w}}\,\dd{\sv{s}},
  \end{equation*}
  where, as before, $\dd{\sv{w}}$ is the counting measure on $\Sp{W}^n$ and $\dd{\sv{s}}$ is Lebesgue measure on $\Sp{S}^n$.
\end{thm}

This is a reflection of the well known result that the Kingman coalescent is projective \citep[][Ch.~3]{Wakeley2008}.
\Cref{thm:synsamp} implies that the genealogy of a synchronous sample contains no information on the population size $n$ unless the event rate $\mu$ is known, and vice versa.
If these parameters are to be independently identifiable, it is necessary to sample asynchronously.

\section{Asynchronous sampling}\label{sec:SMGP}

\subsection{Moran genealogy game with asynchronous sampling}

We now turn to the situation where sampling occurs asynchronously, resulting in a sequence of genealogies whose probabilistic properties we wish to understand.
To represent this, we add some new rules to our parlor game.

\paragraph{Players and equipment}

Before beginning the game, a finite or infinite sequence of \emph{sample times} $0 \le t_1 < t_2 < \dots$ is chosen arbitrarily.
At each of these times, a \emph{single individual} from the population will be sampled.
In addition to the $n$ players of the Moran genealogy game, we must have two players for each of the samples.
The equipment is the same as that used in the Moran genealogy game, with the addition of one red and one blue ball for each sample.
Each one of the two players that will represent a sample receives a slate and a green ball bearing her own (unique) name.
One of these players also takes a blue ball while the other takes a red ball.
The red-ball holders and blue-ball holders are arranged, unseated, in parallel queues.

\paragraph{Setup and play}

The setup for the game with sampling is identical to that for the original game and play proceeds as before.
However, play stops at each of the pre-selected sampling times.
At sampling time $t_k$, the following maneuvers occur (\cref{fig:async_prune}A--B):
\begin{inparaenum}[(1)]
\item Two new seats are placed to the right of the rightmost seat.
\item The number of a randomly chosen black ball is called out.
\item The player holding the corresponding black ball exchanges it for the green ball bearing the name, say A, of the next red-ball holder in the queue.
\item A takes the first of the seats just placed.
\item The next blue-ball holder, call her B, in the queue exchanges her green ball for the red ball held by A.
\item B takes the second of the new seats.
\item Both A and B record the current time ($t_k$) on their slates.
\end{inparaenum}
Thus, after a sampling event, player B sits in the rightmost seat, holding one blue and one red ball.
Player A sits one seat to the left, holding B's green ball and the randomly selected black ball.
An animation depicting a typical simulation of the process can be found in the \href{https://kinglab.eeb.lsa.umich.edu/kingaa/mgp/moran.html#one_simulation_of_the_mgp_with_sampling}{online appendix}.

\subsection{Sampled Moran Genealogy Process}

\paragraph{Definition}

\begin{figure}
\begin{knitrout}\small
\definecolor{shadecolor}{rgb}{0.969, 0.969, 0.969}\color{fgcolor}

{\centering \includegraphics[width=0.95\linewidth]{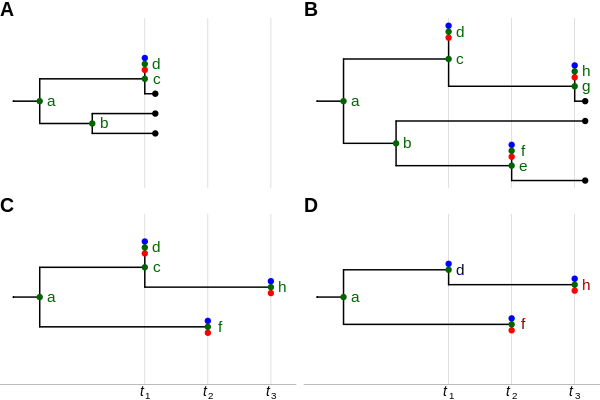} 

}

\end{knitrout}
  \caption{
    \textbf{Sampling and pruning in the Sampled Moran Genealogy Process (SMGP).}
    Here, $n=3$.
    In panel \textbf{A}, we see the configuration shortly after $t_1$, the first sample time.
    Players \textbf{c} and \textbf{d} have just been seated.
    \textbf{c} holds the green ball of \textbf{d} and one black ball;
    \textbf{d} holds one blue and one red ball.
    In panel \textbf{B}, we see the configuration shortly after $t_3$:
    three samples have been taken and players \textbf{e}--\textbf{h} have been seated.
    In this particular realization of the process, no Moran (birth/death) events have occurred in the interval $(t_1,t_3)$.
    One can recover the underlying MGP configuration (not shown) by dismissing first the blue- and then the red-ball holding players.
    Panel \textbf{C} shows the configuration after the black balls have been pruned per \cref{sec:syncsamp}.
    To obtain the observed chain, one additional pruning step must now be applied:
    for every pair of players with identical seating times, one player is dismissed.
    Here, players \textbf{c} and \textbf{d} were both seated at time $t_1$.
    Accordingly, \textbf{c} is dismissed, taking with her one red ball.
    The resulting configuration is depicted in panel \textbf{D}:
    this is $\sg{G}_3$ as defined in the text.
    Player \textbf{d} holds one blue and one green ball:
    he is a \emph{blue player} and corresponds to a \emph{dead sample}.
    Players \textbf{f} and \textbf{h} each hold a red ball:
    they are \emph{red players}, corresponding to \emph{live samples}.
    Finally, there is only one internal node, corresponding to \emph{green player} \textbf{a}.
    The more compact representation of \cref{fig:sample_genealogy} uses a single colored point for each player.
    \label{fig:async_prune}
  }
\end{figure}

The Sampled Moran Genealogy Process (SMGP) is a continuous-time, inhomogeneous, c\`adl\`ag Markov process, $\sg{X}(t)$, on the state space $\Sp{X}\coloneqq\Sp{W}\times\Sp{S}\coloneqq\coprod_{m\in\mathbb{N}}\Sp{W}^m\times\Sp{S}^m$, parameterized by an event rate, $\mu$, a population size $n$, and a finite or infinite sequence of sampling times $\{t_k\}$.
For $t_{k}\le t < t_{k+1}$, we have that $\sg{X}(t)\in\Sp{W}^{n+2k}\times\Sp{S}^{n+2k}$.
The SMGP extends the MGP in the sense that one projects $\sg{X}$ onto $\Sp{X}^n$ by dismissing the players associated with samples.
To be specific, the following procedure projects $\sg{X}(t)$ onto its MGP component (\cref{fig:async_prune}A--B):
\begin{inparaenum}[(i)]
\item Dismiss the players holding blue balls sequentially.
  As usual, a dismissed player trades the ball that is not blue for the ball bearing her name, stands up, and departs.
  The players to the right of the vacated seat each shift one seat to the left, and the rightmost chair is removed.
\item Dismiss every player holding a red ball in the same fashion.
\end{inparaenum}
It is readily verified that at every stage of this procedure, Property G (\cpageref{propertyG}) holds, and the genealogical relationships among the black balls are unchanged.

\paragraph{The observed chain}

We are naturally interested in the genealogies that express the relationships among only the sampled lineages.
Accordingly, we define a mapping $\mathrm{obs}\!:\Sp{X}\to\Sp{X}$ that discards the unobserved lineages.
In particular, for any $t$, let $\mathrm{obs}(\sg{X}(t))$ be the genealogy obtained by performing the following pruning procedure (\cref{fig:async_prune}C--D).
First, sequentially dismiss all players holding black balls, as described in \cref{sec:syncsamp} (\cref{fig:async_prune}C).
Next, each player holding a red ball consults the player immediately to her left.
Let Y be the name of the player with the red ball and X that of the player to the left.
If X and Y were seated at the same time, it is almost surely the case that X holds the green ball bearing Y's name.
In this case, Y trades her red ball for X's other ball.
X now trades the green ball bearing Y's name for the green ball bearing his own name, stands up, and departs.
All players from Y rightward shift one seat to the left and the rightmost seat is removed.
If the slates of X and Y do not match, they take no action.
\Cref{fig:async_prune}D illustrates these maneuvers.
In effect, the pruning procedure strips away structure extraneous to the relationships among samples.

Now, $\mathrm{obs}(\sg{X}(t))$ is constant on each interval $t_k\le t < t_{k+1}$, with jumps at the sample times $t_k$.
Accordingly, the chain $\sg{G}_k\coloneqq\mathrm{obs}(\sg{X}(t_k))$ is well defined.
We refer to $\sg{G}_k$ as the \emph{observed chain} of the SMGP.
Animations showing simulations of the observed chain can be found in the \href{https://kinglab.eeb.lsa.umich.edu/kingaa/mgp/moran.html#simulations_of_the_sampled_moran_genealogy_process}{online appendix}.

Note that, since the genealogies $\sg{G}_{1},\dots,\sg{G}_{k-1}$ are nested within $\sg{G}_k$, the map $\sg{G}_k\mapsto\sg{G}_{j}$ is deterministic for all $j<k$.
This implies that the observed chain, $\sg{G}_k$, is Markov.
Also note that the chain $(\sg{X}(t_k),\sg{G}_k)$ has the structure of a hidden Markov model, also known as a partially observed Markov process \citep{Breto2009,King2016,Smith2017}.
For the remainder of the paper, we will be interested in the probabilistic properties of the observed genealogies $\sg{G}_k$.

\paragraph{Stationarity assumption}

While the SMGP is well defined in the absence of any stationarity assumptions, the theorems below will depend on the stationarity of the underlying MGP.
Accordingly, we will there assume that at some time before the first sample, the state of the SMGP is a random draw from the stationary distribution (\cref{thm:main_mgp}).
We refer to this process as the \emph{stationary SMGP}.

\begin{figure}
\begin{knitrout}\small
\definecolor{shadecolor}{rgb}{0.969, 0.969, 0.969}\color{fgcolor}

{\centering \includegraphics[width=\maxwidth]{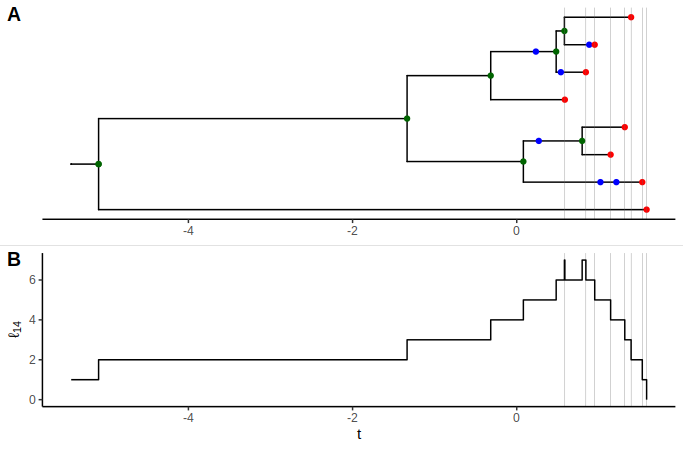} 

}

\end{knitrout}
  \caption{
    \textbf{The observed chain of the Sampled Moran Genealogy Process}.
    \textbf{(A)} The compact tree representation of the $14$-sample genealogy, $\sg{G}_{14}$, from a realization of the stationary SMGP. 
    Green points mark internal nodes; red and blue points indicate samples.
    In terms of the SMGP, each point represents a player of the corresponding color.
    Red and blue players represent live and dead samples, respectively.
    Dead samples correspond to direct-descent events.
    The vertical lines indicate the $e_1,\dots,e_{8}\in\mathrm{live}(\sg{G}_{14})$ as defined in \cref{thm:cond_coal}.
    One can read the attachment times, $a_k$, of each of the samples, from this diagram.
    For example, the attachment time, $a_{14}$, of the 14th sample is that of the leftmost green ball, while $a_{13}$ is that of the rightmost blue ball, an indication that sample 13 descends directly from sample 10.
    Panel \textbf{B} shows the lineage count function $\ell_{14}(t)$, as defined in the text.
    For every $k$, $\ell_k$ is piecewise constant and right continuous.
    It has a unit increase at every green player and a unit decrease at every red player.
    It agrees with the number of lineages in the tree representation of $\sg{G}_{k}$ at all its points of continuity.
    \label{fig:sample_genealogy}
  }
\end{figure}

\paragraph{Direct descent}

In an observed genealogy, it is almost surely the case that no two players have slates that match.
Moreover, each player holding a blue and a red ball corresponds to a sample with no descendants among the other samples;
such a sample is called \emph{live} (\cref{fig:async_prune}D).
On the other hand, each player holding both a blue and a green ball corresponds to a \emph{dead sample}.
A dead sample marks the occurrence of a \emph{direct-descent event}, whereby the lineages of two samples coincide exactly up to the time of the earlier sample (\cref{fig:async_prune}D).

\paragraph{Terminology}

We now define some terms needed in the sequel.
Let $\sg{G}$ be an observed genealogy.
Then $\sg{G}$ is determined by the sequence of seated players, each one of which is characterized by an unordered pair of colored balls (green, red, or blue) and a time.
Let $u_j$ be the time recorded on the slate of the player in seat $j$;
in particular, $u_0=-\infty$.

We can more compactly represent $\sg{G}$ by putting each player into one of three categories:
\emph{green} players are those that hold two green balls;
\emph{blue} players hold one green ball and one blue ball;
\emph{red} players hold one blue and one red ball (cf.~\cref{fig:async_prune,fig:sample_genealogy}).
Green players correspond to branch points in the genealogy;
red players correspond to live samples;
blue players, to dead samples.
Let $\mathrm{live}(\sg{G}_k)=\{u_j:j\in\mathrm{red}(\sg{G}_k)\}$ denote the set of \emph{sample times} of all live samples and let $\mathrm{dead}(\sg{G}_k)=\{u_j:j\in\mathrm{blue}(\sg{G}_k)\}$ be the sample times of dead samples.
Also, let $\mathrm{green}(\sg{G})$ be the \emph{seat numbers} of the green players, with the exception of the player in seat 0.
Similarly, let $\mathrm{blue}(\sg{G})$ and $\mathrm{red}(\sg{G})$ be the seat numbers of the blue and red players, respectively.
Then the number of samples in $\sg{G}$ is $k=\vert\mathrm{blue}(\sg{G})\vert+\vert\mathrm{red}(\sg{G})\vert=\vert\mathrm{live}(\sg{G})\vert+\vert\mathrm{dead}(\sg{G})\vert$.
Moreover, if $r=\vert\mathrm{red}(\sg{G})\vert=\vert\mathrm{live}(\sg{G})\vert$, then $\vert\mathrm{green}(\sg{G})\vert=r-1$ and the total number of seated players is $k+r$.
Note also that $j\in\mathrm{blue}(\sg{G})\,\cup\,\mathrm{red}(\sg{G})$ implies $u_j=t_i$ for some $i$;
the converse holds almost surely.

Now consider the observed chain, $\{\sg{G}_k\}_{k=1}^{\infty}$, with sampling times $\{t_k\}$.
It is clear that $\sg{G}_k$ differs from $\sg{G}_{k-1}$ just in that the lineage of sample $k$ attaches to $\sg{G}_{k-1}$ at some random time $-\infty<A_k\le t_{k-1}$.
This attachment may happen either in a direct-descent event or else at a branching point (\cref{fig:sample_genealogy}A).
If the former, then $\sg{G}_{k}$ differs from $\sg{G}_{k-1}$ in that one red player of $\sg{G}_{k-1}$ has become blue and one red player has been added;
if the latter, then $\sg{G}_{k}$ has added two players (one red and one green) to $\sg{G}_{k-1}$.

Given a realization, $\{\sg{G}_k\}$, of the observed chain of the SMGP, one can unambiguously define the \emph{attachment times}, $\{a_k\}_{k=2}^{\infty}$ so that $a_k$ is the time at which the lineage of sample $k$ attaches to $\sg{G}_{k-1}$.
Note that $a_1$ is undefined and that $a_k\le t_{k-1}<t_k$ for $k>1$.
Notice also that every green player corresponds to an attachment:
$j\in\mathrm{green}(\sg{G}_k)$ when, and only when, $u_j=a_i$ for some $i$.
Likewise, every blue player corresponds to both a sample and an attachment:
$j\in\mathrm{blue}(\sg{G}_k)$ if and only if there are $i_1$ and $i_2$, $i_1<i_2$, such that $u_j=t_{i_1}=a_{i_2}$.

Define the lineage-count function $\ell_k:\Sp{R}\to\Sp{N}$ so that, for every $t$, $\ell_k(t)$ is the number of live samples in $\sg{G}_{k}$ with seating times greater than $t$ minus the number of branch points with times greater than $t$.
That is
\begin{equation}\label{eq:ell}
  \ell_k(t)\;\coloneqq\;\left\vert\left\{j\in\mathrm{red}(\sg{G}_k):u_j>t\right\}\right\vert-\left\vert\left\{j\in\mathrm{green}(\sg{G}_k):u_j>t\right\}\right\vert.
\end{equation}
With this definition, $\ell_k$ is right continuous with left limits (c\`adl\`ag).
In particular, $e\in\mathrm{live}(\sg{G}_k)$ implies that $\ell_k$ decreases, almost surely, by one unit at time $e$.
By contrast, $\ell_k$ is continuous almost surely at $e\in\mathrm{dead}(\sg{G}_k)$.
In terms of the tree representation of $\sg{G}_{k}$  (\cref{fig:sample_genealogy}B), $\ell_k(t)$ is the number of lineages at time $t$ wherever the latter is unambiguous.
Note that $\ell_k(t)=1$ for $t<u_1$ and $\ell_k(t)=0$ for $t\ge t_{k}$.

\subsection{Transition probabilities}

We are interested in the probability measure on observed genealogies, as generated by the stationary SMGP.
To obtain this, we will begin by deriving an expression for the measure of $\sg{G}_{k+1}$ conditional on $\sg{G}_{k}$.
\Cref{thm:cond_coal} depends on \cref{lemma:cumhaz,lemma:nugget}, the statements of which we temporarily postpone.

\begin{thm}\label{thm:cond_coal}
  Let $\sg{G}_k$ be the observed chain of the stationary SMGP and let $\mathrm{live}(\sg{G}_k)=\left\{e_1, \dots, e_q\right\}$, where $e_1 < e_2 < \cdots < e_q$.
  Then, we have the following
  \begin{align}
    -\log\prob{A_{k+1}<a\;\vert\;\sg{G}_{k}}\;&=\;\int_a^{\infty}\!\frac{\ell_k(t)}{\binom{n}{2}}\,\mu\,\dd{t}\;+\;\sum_{\mathclap{\left\{j\;:\;a\,\le\,e_j\right\}}}\,\log{\frac{n-\ell_k(e_j)}{n-\ell_k(e_j)-1}},\label{eq:cond_coal_cdf_strict}\\
    -\log\prob{A_{k+1}\le a\;\vert\;\sg{G}_{k}}\;&=\;\int_a^{\infty}\!\frac{\ell_k(t)}{\binom{n}{2}}\,\mu\,\dd{t}\;+\;\sum_{\mathclap{\left\{j\;:\;a\,<\,e_j\right\}}}\,\log{\frac{n-\ell_k(e_j)}{n-\ell_k(e_j)-1}},\label{eq:cond_coal_cdf}\\
    -\log\prob{A_{k+1}=a\;\vert\;\sg{G}_{k}}\;&=\;\begin{cases}
    \log{(n-\ell_k(a))}-\log\prob{A_{k+1}\le a}, &a\in\mathrm{live}(\sg{G}_k),\\
    0, &a\notin\mathrm{live}(\sg{G}_k).
    \end{cases}\label{eq:cond_coal_discrete}
  \end{align}
  Moreover, the probability density of $A_{k+1}$, conditional on $\sg{G}_k$, is given by
  \begin{equation}\label{eq:cond_coal_pdf}
    f_{A_{k+1}\vert\sg{G}_k}(a)\,\dd{a}=\prob{A_{k+1}\le a\;\vert\;\sg{G}_{k}}\,\left(\mu\,\frac{\ell_k(a)}{\binom{n}{2}}\,\dd{a}+\frac{\indicator{a\in\mathrm{live}(\sg{G}_k)}}{n-\ell_k(a)}\,\dd{n}\right),
  \end{equation}
  where $\dd{a}$ signifies Lebesgue measure and $\dd{n}$, counting measure, both on $\Sp{R}$.
  \begin{proof}
    Let $J(a)\coloneqq\min\{j:a\le e_j\}$.
    It is an identity that
    \begin{equation*}
      \begin{split}
        \prob{A_{k+1}<a}\;=\;&\prob{A_{k+1}<a\vert A_{k+1}<e_{J(a)}}\,\times\,\prod_{j=J(a)}^{q}\!\prob{A_{k+1}<e_{j}\vert A_{k+1}\le e_{j}}\\
        &\times\,\prod_{j=J(a)}^{q-1}\!\prob{A_{k+1}\le e_{j}\vert A_{k+1}< e_{j+1}}\,\times\,\prob{A_{k+1}\le e_q}.
      \end{split}
    \end{equation*}
    Now, by \cref{lemma:cumhaz},
    \begin{equation*}
      \prob{A_{k+1}<a\vert A_{k+1}<e_{J(a)}}=\exp{\left(-\int_{a}^{e_{J(a)}}\!\frac{\ell_k(t)}{\binom{n}{2}}\,\mu\,\dd{t}\right)}
    \end{equation*}
    and also, for every $j$,
    \begin{equation*}
      \prob{A_{k+1}\le e_{j}\vert A_{k+1}< e_{j+1}}=\exp{\left(-\int_{e_{j}}^{e_{j+1}}\!\frac{\ell_k(t)}{\binom{n}{2}}\,\mu\,\dd{t}\right)}.
    \end{equation*}
    On the other hand, by \cref{lemma:nugget}, we have
    \begin{equation*}
      \prob{A_{k+1}<e_{j}\vert A_{k+1}\le e_{j}} = 1-\prob{A_{k+1}=e_{j}\vert A_{k+1}\le e_{j}}=\frac{n-\ell_k(e_{j})-1}{n-\ell_k(e_{j})},
    \end{equation*}
    for all $j$.
    Finally, note that $e_{q}=t_{k}$ and that, therefore, $\prob{A_{k+1}\le e_q}=1$.
    Putting these all together, we obtain \cref{eq:cond_coal_cdf_strict}, with \Cref{eq:cond_coal_cdf,eq:cond_coal_discrete,eq:cond_coal_pdf} as elementary consequences.
  \end{proof}
\end{thm}



\begin{lemma}\label{lemma:cumhaz}
  With the definitions as in \cref{thm:cond_coal},
  \begin{equation*}
    \prob{A_{k+1}<a\vert A_{k+1}<e_{j}}=\exp{\left(-\int_a^{e_{j}}\!\frac{\ell_k(t)}{\binom{n}{2}}\,\mu\,\dd{t}\right)},\ \text{whenever}\ e_{j-1}<a<e_{j}.
  \end{equation*}
  Moreover,
  \begin{equation*}
    \prob{A_{k+1}\le e_{j-1}\vert A_{k+1}<e_{j}}=\exp{\left(-\int_{e_{j-1}}^{e_{j}}\!\frac{\ell_k(t)}{\binom{n}{2}}\,\mu\,\dd{t}\right)}.
  \end{equation*}
  \begin{proof}
    Viewing the lineage attachment as a survival process in backward time, it is sufficient to show that the hazard of attachment is
    \begin{equation*}
      \lambda(t) = \frac{\ell_k(t)}{\binom{n}{2}}\,\mu.
    \end{equation*}
    To see this, let $a$ and $\eps>0$ be such that $e_{j-1}\le a<a+\eps<\min\{t_i:t_i>e_{j-1}\}$.
    Note that, since the interval $(a,a+\eps)$ lies between adjacent sample-times, no direct-descent events can have occurred in this interval.
    Therefore, conditional on $A_{k+1}<a+\eps$, coalescence of the lineage of sample $k+1$ with $\sg{G}_{k}$ occurs within this interval if and only if (1) a birth event occurs in the interval and (2) the associated parent/child pair includes the unique ancestor of sample $k+1$ and one of the $\ell_k(a)=\ell_k(e_{j-1})$ players ancestral to the first $k$ samples.
    The probability that a Moran event occurred in the interval $(a,a+\eps)$ is $\mu\,\eps+o(\eps)$.
    Conditional on $A_{k+1}<a+\eps$, the unique ancestor of sample $k+1$ at time $a+\eps$, by definition, is not among the $\ell_k(a)=\ell_k(e_{j-1})$ lineages of $\sg{G}_{k}$ present at this time.
    Therefore, if a Moran event has occurred in the interval, of the $\binom{n}{2}$ pairs that might have been involved in the event, exactly $\ell_k(a)$ of these involve one of the lineages of $\sg{G}_{k}$ and the unique ancestor of sample $k+1$.
    Since all of these pairs are equally likely to have been involved, the probability that a coalescence event occurs in the interval is
    \begin{equation*}
      \prob{a<A_{k+1}\vert A_{k+1}<a+\eps}=\frac{\ell_k(a)}{\binom{n}{2}}\,\mu\,\eps+o(\eps)=\lambda(a)\,\eps+o(\eps).
    \end{equation*}
    Finally, note that, if $t_i\in(e_{j-1},e_j)$, then $t_i\in\mathrm{dead}(\sg{G}_k)$, whence $\prob{A_{k+1}=t_i}=0$.
    The second equation in the statement of the lemma follows from the fact that $\ell_k$ is right continuous.
  \end{proof}
\end{lemma}

\begin{lemma}\label{lemma:nugget}
  With the definitions as in \cref{thm:cond_coal}, we have
  \begin{equation*}
    \prob{A_{k+1}=e_j\vert A_{k+1}\le e_j} = \frac{1}{n-\ell_k(e_j)}.
  \end{equation*}
  \begin{proof}
    If $A_{k+1}\le e_j$, then by definition, the unique ancestor of sample $k+1$ at time $e_j$ cannot be any one of the $\ell_k(e_j)$ individuals ancestral at time $e_j$ to the first $k$ samples.
    However, it is equally likely to be any one of the $n-\ell_k(e_j)$ other members of the population.
    Of these, exactly one corresponds to the sample at $e_j$.
  \end{proof}
\end{lemma}

\Cref{thm:cond_coal} establishes the probability distribution of $A_k\vert\sg{G}_{k-1}$;
it is only a short step to that of $\sg{G}_{k}\vert\sg{G}_{k-1}$.
Let $f_{\sg{G}_k\vert\sg{G}_{k-1}}(a)$ denote the probability density function of $\sg{G}_{k}$ conditional on $\sg{G}_{k-1}$, evaluated at attachment time $a_k=a$.

\begin{corol}\label{thm:cond_genealogy}
  The conditional probability density of $\sg{G}_k\vert\sg{G}_{k-1}$ is
  \begin{equation*}
    f_{\sg{G}_{k}\vert\sg{G}_{k-1}}(a_k)\,\dd{a_k}=\prob{A_{k}\le a_k\;\vert\;\sg{G}_{k-1}}\,\left(\frac{\mu}{\binom{n}{2}}\,\dd{a_k}+\frac{\indicator{a_k\in\mathrm{live}(\sg{G}_{k-1})}}{n-\ell_{k-1}(a_k)}\,\dd{n_k}\right),
  \end{equation*}
  where $\dd{a_k}$ and $\dd{n_k}$ are, respectively, Lebesgue and counting measure on $\Sp{R}$, the space of allowable attachment times $a_k$.
  \begin{proof}
    By \cref{thm:cond_coal}, we have that
    \begin{equation}\label{eq:smgp_cond_coal}
      f_{A_{k}\vert\sg{G}_{k-1}}(a_k)\,\dd{a_k}=\prob{A_{k}\le a_k\;\vert\;\sg{G}_{k-1}}\,\left(\mu\,\frac{\ell_{k-1}(a_k)}{\binom{n}{2}}\,\dd{a_k}+\frac{\indicator{a_k\in\mathrm{live}(\sg{G}_{k-1})}}{n-\ell_{k-1}(a_k)}\,\dd{n_k}\right).
    \end{equation}
    The second factor in \cref{eq:smgp_cond_coal} has two terms, the first of which accounts for the attachment of the $k$-th sample lineage in one of the intervals between two players of $\sg{G}_{k-1}$.
    When such an attachment occurs, there are precisely $\ell_{k-1}(a_k)$ lineages in $\sg{G}_{k-1}$ to which the new lineage might attach.
    Equivalently, there are $\ell_{k-1}(a_k)$ green balls bearing the names of players to the right of $a_k$ held by players to the left, one of which is selected at random upon attachment.
    Under the assumption that the underlying MGP is stationary, each of these is equally likely.
    On the other hand, when the new lineage attaches via a direct-descent event, there is (almost surely) only one choice as to where the attachment will occur.
  \end{proof}
\end{corol}

\subsection{Marginal distribution of observed genealogies}

\Cref{thm:cond_genealogy} establishes the probability distribution of each $\sg{G}_k$, conditional on $\sg{G}_{k-1}$.
We can use this to compute the probability distribution for any sequence of genealogies, $\left\{\sg{G}_j\right\}_{j=1}^k$ generated by the stationary SMGP.
In particular, we will derive expressions for probability measures on the space of genealogies $\Sp{X}$.
When the underlying MGP is stationary, these will all be uniform with respect to the genealogies' discrete aspect (the sequence of pairs of colored balls), but will have nontrivial dependence on the continuous aspect, i.e., the attachment times $\{a_k\}_{k=2}^{\infty}$ of the second and successive samples.
Accordingly, we will focus on the latter.
Specifically, we will denote the probability measure on the space of $k$-sample observed genealogies by
\begin{equation*}
  f_{\sg{G}_k}(\sv{a})\,\dd{\sv{a}}
\end{equation*}
where $\sv{a}=(a_2,\dots,a_k)$ is the vector of attachment times, $\dd{\sv{a}}=\dd{a_2}\,\cdots\,\dd{a_k}$ denotes Lebesgue measure on $\Sp{R}^{k-1}$, and $f_{\sg{G}_k}$ is a probability density function.

We begin by establishing some elementary properties of the lineage-count functions, $\ell_k$, defined above.

\begin{lemma}\label{lemma:continuous_part}
  Let $\left\{\sg{G}_k\right\}_{k=1}^\infty$ be the observed chain of the SMGP, with sample times $\{t_k\}$.
  Let $\{a_k\}_{k=2}^{\infty}$ be the attachment times of each of the successive samples.
  Then
  \begin{equation*}
    \sum_{j=2}^k\int_{a_j}^{\infty}\!\ell_{j-1}(t)\,\dd{t} = \int_{-\infty}^{\infty}\!\binom{\ell_k(t)}{2}\,\dd{t}.
  \end{equation*}
  \begin{proof}
    We argue by induction on $k$.
    First, note that $a_2\le t_1<t_2$.
    Moreover,
    \begin{equation*}
      \begin{gathered}
        \ell_1(t)=\begin{cases}
        1, &t<t_1,\\
        0, &\text{otherwise},
        \end{cases}\qquad
        \text{and}\qquad
        \ell_2(t)=\begin{cases}
        1, &t<a_2,\\
        2, &a_2\le t<t_1,\\
        1, &t_1\le t<t_2,\\
        0, &\text{otherwise}.
        \end{cases}
      \end{gathered}
    \end{equation*}
    It follows that
    \begin{equation*}
      \int_{a_2}^{\infty}\!\ell_1(t)\,\dd{t}=t_1-a_2=\int_{a_2}^{t_1}\!\dd{t}=\int_{-\infty}^{\infty}\!\binom{\ell_2(t)}{2}\,\dd{t}.
    \end{equation*}
    Now, we suppose that the result holds for $k$ and observe that this implies
    \begin{equation*}
      \begin{split}
        \sum_{j=2}^{k+1}\int_{a_j}^{\infty}\!\ell_{j-1}(t)\,\dd{t}&=\int_{-\infty}^{\infty}\!\binom{\ell_k(t)}{2}\,\dd{t}\;+\;\int_{a_{k+1}}^{\infty}\!\ell_{k}(t)\,\dd{t}\\
        &= \int_{-\infty}^{a_{k+1}}\!\binom{\ell_k(t)}{2}\,\dd{t}+\int_{a_{k+1}}^{t_{k+1}}\!\binom{\ell_{k}(t)+1}{2}\,\dd{t}+\int_{t_{k+1}}^{\infty}\!\binom{\ell_k(t)+1}{2}\,\dd{t}\\
        &= \int_{-\infty}^{\infty}\!\binom{\ell_{k+1}(t)}{2}\,\dd{t}.
      \end{split}
    \end{equation*}
    Here, we have used the identity $\binom{m}{2}+m=\binom{m+1}{2}$ and the facts that $\ell_k(t)=0$ for $t>t_{k+1}$ and
    \begin{equation*}
      \ell_{k+1}(t)=\begin{cases}
      \ell_{k}(t)+1, &a_{k+1}\le t<t_{k+1},\\
      \ell_{k}(t),   &\text{otherwise}.
      \end{cases}
    \end{equation*}
  \end{proof}
\end{lemma}

Now observe that sample $j$, taken at time $t_j$, is live in $\sg{G}_j$.
With each subsequent sample, there is a chance that it will die.
While it remains alive, however, each subsequent sample may attach to the left or to the right of $t_j$.
Define $m(j,k)$ to be the number of samples that attach to the left of $t_j$ up to the point that sample $j$ dies or $k$ is reached.
That is, $m(j,k)=\left\vert\left\{i:j<i\le k\ \text{and}\ a_{i}<t_{j}\ \text{and}\ a_{r}\ne t_{j}\ \text{for all}\ r<i\right\}\right\vert$.
When $t_j\in\mathrm{dead}(\sg{G}_k)$, then $m(j,k)$ is the lineage count at $t_j$ at the time when $t_j$ was killed, i.e., $a_{i}=t_j$ implies $m(j,k)=\ell_{i-1}(t_j)$.
Likewise, $t_j\in\mathrm{live}(\sg{G}_k)$ implies $m(j,k)=\ell_k(t_j)$.

\begin{lemma}\label{lemma:discrete_part}
  Let $\left\{\sg{G}_k\right\}_{k=1}^\infty$ be the observed chain of the SMGP, with sample times $\{t_k\}_{k=1}^{\infty}$ and attachment times $\{a_k\}_{k=2}^{\infty}$.
  Define $m(j,k)$ as above.
  Then
  \begin{equation*}
    \sum_{j=1}^{k-1}\sum_{{e\in\mathrm{live}(\sg{G}_{j})}}\log{\frac{n-\ell_{j}(e)}{n-\ell_{j}(e)-1}}\,\indicator{e>a_{j+1}}\;=\;\sum_{j=1}^{k-1}\log{\frac{n}{n-m(j,k)}}.
  \end{equation*}
  \begin{proof}
    \begin{equation*}
      \begin{split}
        S\;\coloneqq\;&\sum_{j=1}^{k-1}\sum_{{e\in\mathrm{live}(\sg{G}_{j})}}\log{\frac{n-\ell_{j}(e)}{n-\ell_{j}(e)-1}}\,\indicator{e>a_{j+1}}\;=\;
        \sum_{j=1}^{k-1}\sum_{i=1}^{j}\indicator{t_{i}\in\mathrm{live}(\sg{G}_{j})}\,\indicator{t_{i}>a_{j+1}}\,\log{\frac{n-\ell_{j}(t_{i})}{n-\ell_{j}(t_{i})-1}}\\
        =&\sum_{i=1}^{k-1}\sum_{j=i}^{k-1}\indicator{t_{i}\in\mathrm{live}(\sg{G}_{j})}\,\indicator{t_{i}>a_{j+1}}\,\log{\frac{n-\ell_{j}(t_{i})}{n-\ell_{j}(t_{i})-1}}.
      \end{split}
    \end{equation*}
    Now, note that $t_{i}\in\mathrm{live}(\sg{G}_j)$ and $t_{i}>a_{j+1}$ if and only if $t_{i}\in\mathrm{live}(\sg{G}_{j+1})$ and $t_{i}>a_{j+1}$.
    Therefore,
    \begin{equation*}
      S = \sum_{i=1}^{k-1}\sum_{j=i+1}^{k}\indicator{t_{i}\in\mathrm{live}(\sg{G}_{j})}\,\indicator{t_{i}>a_{j}}\,\log{\frac{n-\ell_{j-1}(t_{i})}{n-\ell_{j-1}(t_{i})-1}}.
    \end{equation*}
    The inner sum contains $m(i,k)$ terms:
    one for each sample $j>i$ such that $a_{j}<t_{i}$ up to the sample (if any) for which $a_{j}=t_{i}$, at which point sample $i$ dies.
    Because for each such sample $j$, $\ell_{j}(t_{i})=\ell_{j-1}(t_{i})+1$, we have
    \begin{equation*}
      S = \sum_{i=1}^{k-1}\,\sum_{j=0}^{m(i,k)-1}\log{\frac{n-\ell_{i}(t_i)-j}{n-\ell_{i}(t_i)-j-1}} = \sum_{i=1}^{k-1}\log{\frac{n}{n-m(i,k)}},
    \end{equation*}
    where in the last equation, we have used the fact that $\ell_i(t_i)=0$ for all $i$.
  \end{proof}
\end{lemma}

We can now state the main result of this section, which gives the joint probability distribution of any sequence of observed genealogies and, equivalently, the unconditional probability distribution of each observed genealogy generated by the stationary SMGP.

\begin{thm}\label{thm:smgp_dist}
  Let $\left\{\sg{G}_k\right\}_{k=1}^{\infty}$ be the observed chain of the stationary SMGP, with sampling times $\{t_k\}$ and attachment times $\{a_k\}$.
  Then
  \begin{equation}\label{eq:smgp_dist}
    \begin{split}
      f_{\sg{G}_1,\dots,\sg{G}_k}(\sv{a})\,\dd{\sv{a}}=&n^{r-k}\,\left(\frac{\mu}{\smash{\binom{n}{2}}}\right)^{r-1}\,\exp{\left(-\int_{-\infty}^{\infty}\!\mu\,\frac{\binom{\ell_k(t)}{2}}{\binom{n}{2}}\,\dd{t}\right)}\\
      &\times\prod_{\{i:\nexists j>i\ a_j=t_i\}}\left(1-\frac{\ell_k(t_i)}{n}\right)\,\prod_{\{j:\nexists i<j\ a_j=t_i\}}\dd{a_j}\,\prod_{\{j:\exists i<j\ a_j=t_i\}}\dd{n_j}.
    \end{split}
  \end{equation}
  where $r=\vert\mathrm{live}(\sg{G}_k)\vert=\vert\{j:\nexists i\ a_j=t_i\}\vert$.
  Moreover, $f_{\sg{G}_1,\dots,\sg{G}_k}=f_{\sg{G}_k}$.
  \begin{proof}
    The joint probability density of $\left\{\sg{G}_j\right\}_{j=1}^{k}$ is the product of the one-step conditional probability densities:
    \begin{equation}\label{eq:smgp_dist_A}
      \begin{split}
        f_{\sg{G}_1,\dots,\sg{G}_k}(\sv{a})\,\dd{\sv{a}}&=\prod_{j=2}^{k}f_{\sg{G}_j\vert\sg{G}_{j-1}}(a_j)\,\dd{a_j}\\
        &=\prod_{j=2}^{k}\prob{A_j\le a_j\;\vert\;\sg{G}_{j-1}}\,\prod_{j=2}^{k}\left(\frac{\mu}{\binom{n}{2}}\,\dd{a_j}+\frac{\indicator{a_j\in\mathrm{live}(\sg{G}_{j-1})}}{n-\ell_{j-1}(a_j)}\,\dd{n_j}\right).
      \end{split}
    \end{equation}
    Let $F$ denote the first product in the last expression and $G$, the second.
    By \cref{lemma:continuous_part,lemma:discrete_part}, we can simplify $F$:
    \begin{equation*}\label{eq:smgp_dist_B}
      F=\exp{\left(-\int_{-\infty}^{\infty}\!\mu\,\frac{\binom{\ell_k(t)}{2}}{\binom{n}{2}}\,\dd{t}\right)}\,\prod_{j=1}^{k}\frac{n-m(j,k)}{n}.
    \end{equation*}
    Here we have used the fact that $m(k,k)=0$.
    Now, $G$ contains one factor for each sample.
    We can divide it into two sub-products, according to whether the sample was a direct descendant of an earlier sample or not:
    \begin{equation}\label{eq:smgp_dist_B1}
      G=\prod_{\{j:a_j\notin\mathrm{live}(\sg{G}_{j-1})\}}\frac{\mu}{\binom{n}{2}}\,\dd{a_j}\,\prod_{\{j:a_j\in\mathrm{live}(\sg{G}_{j-1})\}}\frac{\dd{n_j}}{n-\ell_{j-1}(a_j)}.
    \end{equation}
    In passing from \cref{eq:smgp_dist_A} to \cref{eq:smgp_dist_B1}, we have discarded terms proportional to $\indicator{a_j\in\mathrm{live}(\sg{G}_{j-1})}\dd{a_i}\,\dd{n_j}$ which vanish off a set of Lebesgue measure zero.
    Now we notice that the first product in \cref{eq:smgp_dist_B1} has one factor for each green player in $\sg{G}_k$, while the second product has one factor for each blue player.
    For $j\in\mathrm{green}(\sg{G}_k)\cup\mathrm{blue}(\sg{G}_k)$, let $s(j)$ the number of the unique sample that attaches at $j$, i.e., $u_j=a_{s(j)}$.
    With this definition, we have
    \begin{equation}\label{eq:smgp_dist_C}
      G=\prod_{j\in\mathrm{green}(\sg{G}_k)}\frac{\mu}{\binom{n}{2}}\,\dd{a_{s(j)}}\,\prod_{j\in\mathrm{blue}(\sg{G}_k)}\frac{\dd{n_{s(j)}}}{n-m(s(j),k)}.
    \end{equation}
    Since every sample is either a red or a blue player, \Cref{eq:smgp_dist_C} is equivalent to
    \begin{equation*}\label{eq:smgp_dist_D}
      \begin{split}
        G&=\left(\frac{\mu}{\smash{\binom{n}{2}}}\right)^{r-1}\,\prod_{j=1}^{k}\frac{1}{n-m(j,k)}\,\prod_{e\in\mathrm{live}(\sg{G}_k)}\left(n-\ell_k(e)\right)\,\prod_{{j\in\mathrm{green}(\sg{G}_k)}}\dd{a_{s(j)}}\,\prod_{{j\in\mathrm{blue}(\sg{G}_k)}}\dd{n_{s(j)}}\\
        &=\left(\frac{\mu}{\smash{\binom{n}{2}}}\right)^{r-1}\,\prod_{j=1}^{k}\frac{1}{n-m(j,k)}\,\prod_{\{i:\nexists j>i\ a_j=t_i\}}\left(n-\ell_k(t_i)\right)\,\prod_{\{j:\nexists i<j\ a_j=t_i\}}\dd{a_j}\,\prod_{\{j:\exists i<j\ a_j=t_i\}}\dd{n_j},\\
      \end{split}
    \end{equation*}
    where $r=\vert\{j:\nexists i\ a_j=t_i\}\vert$ is the number of red players (live samples) in $\sg{G}_k$,
    Returning to \cref{eq:smgp_dist_A}, we obtain \cref{eq:smgp_dist}.
    Finally, the last statement in the theorem follows trivially in view of the fact that, for each $k$, $\prob{\sg{G}_{k-1}\;\vert\;\sg{G}_k}=1$.
  \end{proof}
\end{thm}

The log likelihood is of great importance from an inference point of view.
It is given explicitly in the following
\begin{corol}\label{thm:loglik}
  For $k>1$, if $\sg{G}_k$ is a $k$-sample genealogy drawn from the observed chain of the stationary Sampled Moran Genealogy Process with population size $n$, event rate $\mu$, sampling times $t_1,\dots,t_k$, and attachment times $a_2,\dots,a_k$, then the log likelihood is
  \begin{equation*}
    \log\mathcal{L}=(r-k)\,\log{n}+(r-1)\,\log\frac{\mu}{\smash{\binom{n}{2}}}-\int_{-\infty}^{\infty}\!\frac{\binom{\ell_k(t)}{2}}{\binom{n}{2}}\,\mu\,\dd{t}+\sum_{\mathclap{e\in\mathrm{live}(\sg{G}_k)}}\log{\left(1-\frac{\ell_k(e)}{n}\right)},
  \end{equation*}
  where $r=\vert\mathrm{live}(\sg{G}_k)\vert$ and $\ell_k$ is defined by \cref{eq:ell}.
\end{corol}
In view of the form of the log likelihood given by \cref{thm:loglik}, it is clear that the population size $n$ and event rate $\mu$ are individually identifiable on the basis of sequentially sampled genealogies.

\section{Discussion}\label{sec:discussion}

The recent paper by \citet{Wirtz2019} defines the \emph{Evolving Moran Genealogy} Markov chain, which is identical to our $\sv{W}$ chain (which encodes the dynamics of the topological structure of the genealogies, ignoring branch lengths) when the latter is stationary.
These authors establish a number of results regarding this process, for finite population sizes, including derivations of the evolution of tree balance statistics and the form of the process' time-reversal.

\citet{Etheridge2019} extend the look-down construction of \citet{Donnelly1996,Donnelly1999} to a much richer class of demographies than we consider here:
Moran demography is only one of the simpler special cases their elegant abstract approach subsumes.
However, \citet{Etheridge2019} are principally concerned with deriving results in the infinite-population limit.
Nor do they consider the effects of asynchrous sampling or direct descent, as we do here.

If the sampling times $t_k$ are a Poisson process with rate $\nu$, and if $\nu$ is much smaller than the Moran event rate $\mu$, one will have $r\sim k$ and $\ell_k\ll n$ in \cref{thm:loglik}.
In this case, we have following approximation to the log likelihood:
\begin{equation}\label{eq:ll_approx}
  \log\mathcal{L}\approx (k-1)\,\log\frac{\mu}{\smash{\binom{n}{2}}}-\int_{-\infty}^{\infty}\!\frac{\binom{\ell_k(t)}{2}}{\binom{n}{2}}\,\mu\,\dd{t}-\sum_{j=1}^k\frac{\ell_k(t_j)}{n}.
\end{equation}
One can compare this quantity with that obtained from specializing the phylodynamic methods of \citet{Volz2009a} and \citet{Rasmussen2011} to the case of Moran demography.
In the same limit ($\nu\ll\mu$) and with $n\to\infty$, the latter methods agree and give an expression for the likelihood of a given genealogy that, in our notation, is
\begin{equation}\label{eq:vr_ll}
  \log\mathcal{L}_{\mathrm{VRK}}=(k-1)\,\log\frac{\mu}{\smash{\binom{n}{2}}}-\int_{-\infty}^{\infty}\!\frac{\binom{\ell_k(t)}{2}}{\binom{n}{2}}\,\mu\,\dd{t}+\sum_{\mathclap{i\in\mathrm{green}(\sg{G}_k)}}\log\binom{\ell_k(a_i)}{2}.
\end{equation}
Comparing \cref{eq:ll_approx,eq:vr_ll}, we see that the expressions differ by two terms, one of which depends only on the data and is therefore irrelevant from the perspective of inference.
The term that remains is of order $\frac{k}{n}\,\langle\ell_k\rangle$, where $\langle\ell_k\rangle$ is the mean of $\ell_k(t)$ across sampling times.
Since $\langle\ell_k\rangle\sim n\,\sqrt{{\nu}/{\mu}}$ as $n\to\infty$ when $\nu\ll\mu$, we see that this discrepancy is roughly $\sqrt{{\nu}/{\mu}}$ per sample.
Thus the expressions of \citet{Volz2009a} and \citet{Rasmussen2011} are good approximations when sampling is relatively sparse and population sizes are large.

More generally, the method of \citet{Rasmussen2011} was derived using layers of approximations that we have shown to be unnecessary.
In particular, \citet{Volz2009a} derived a coalescent likelihood in a large-population deterministic limit;
\citet{Rasmussen2011} then used this as an approximate likelihood for a stochastic model.
By contrast, we have derived an exact formula similar to that of \citet{Volz2009a} but which applies to a stochastic dynamic model for all population and sample sizes.

A significant achievement of \citet{Volz2009a} was to improve on previous attempts to apply coalescent methods for time-varying populations.
The present paper does not directly address this extension, but it has not escaped our notice that analogues of \cref{lemma:cumhaz,lemma:nugget}, and therefore of \cref{thm:cond_coal,thm:smgp_dist}, exist for a broad class of birth-death processes, though generalization of these results is beyond the scope of the present paper.
In future work, we will develop exact inference methodology that improves upon the heuristic proposal of \citet{Rasmussen2011}.

\section*{Acknowledgements}

The authors gratefully acknowledge useful conversations with Simon Frost, David Rasmussen, Jonathan Terhorst, Mitchell Newberry, and two anonymous reviewers.
This work was supported by grants from the U.S. National Institutes of Health, (Grant \#1R01AI143852 to AAK, \#1U54GM111274 to AAK and ELI) and a grant from the Interface program, jointly operated by the U.S. National Science Foundation and the National Institutes of Health (Grant \#1761603 to ELI and AAK).
QL was supported by a fellowship from the Michigan Institute for Data Science.

\section*{Online Appendix}

An \href{https://kinglab.eeb.lsa.umich.edu/kingaa/mgp/moran.html}{online appendix} is available, containing illustrative animations, simulator code, and numerical verification of some of the statements proved in the text.
These materials will be permanently archived upon acceptance of the paper.

\bibliographystyle{preprint}
\bibliography{phylopomp}

\appendix


\section{Properties of the MGP}\label{sec:properties}

In this appendix, we derive a number of basic results about the Moran Genealogy Process, as represented by the Game.
Almost all of the results in this section are well known, and none of them will be surprising to the reader familiar with the literature on the Moran process and the Kingman coalescent.
Nevertheless, some of the proofs given here are novel, as will be indicated below.
In particular, the fact that they are purely forward-looking with respect to time may be of of some interest.

\subsection{Cardinality of $\Sp{W}^n$}
\label{sec:w-size}

It is clear that the specific names of the genealogy game players are irrelevant, provided they are unique.
Accordingly, we ignore the identities of the players entirely in the counting.
One can readily count the number of distinct arrangements to compute the size of $\Sp{W}^n$.
We see that there are $\binom{n}{2}$ choices for the two black balls held by the player in seat $n-1$.
The player in seat $n-2$ now has $n-1$ balls to choose from:
the remaining $n-2$ black balls plus the green ball with the name of the player in seat $n-1$.
Continuing to work backward, for each $m$, the player in seat $m$ has $\binom{m+1}{2}$ choices for a pair of balls.
Hence, one has $\vert\Sp{W}^n\vert=\prod_{m=1}^{n-1}\!\binom{m+1}{2}=n!\,(n-1)!/2^{n-1}$.

\subsection{Limiting distribution of $\sv{W}$}
\label{sec:w-dist}

With the definitions of \cref{sec:MGP}, it should be clear that both $\sv{X}$ and $\sv{W}$ are Markov processes, though $\sv{S}$ is not.
Here, we show that the unique, limiting distribution of the $\sv{W}$ chain is the uniform distribution on $\Sp{W}^n$.
This well known fact has been proved by many authors \citep[e.g.,][]{Aldous1999,Gernhard2008,Wirtz2019}.
For completeness and to introduce notation, we provide a proof here, which closely follows the reasoning of \citet{Aldous1999}.

\begin{prop}
  \label{prop:q}
  Let $q_n$ be the uniform probability distribution on $\Sp{W}^n$.
  That is, for each $\sv{w}\in\Sp{W}^n$,
  \begin{equation*}
    q_n(\sv{w}) \coloneqq \frac{2^{n-1}}{n!\,(n-1)!}.
  \end{equation*}
  Then $q_n$ is the unique, limiting, stationary distribution of the $\sv{W}$ process described above.
\end{prop}
\begin{proof}
  It is easy to verify that $\sv{W}$ is irreducible, aperiodic, and recurrent, whence it follows that it has a unique, limiting invariant distribution.
  The key to establishing that this distribution is uniform is to retrace the steps of the last-seated player, counting the number of states that can immediately precede a given state.

  For $\sv{w}\in\Sp{W}^n$, let $\Upsilon_u(\sv{w})\in\Sp{W}^{n-1}$ be the configuration resulting from the ``killing'' of the player who holds ball $u$.
  For $\sv{w}\in\Sp{W}^{n-1}$, let $\Phi_{v}(\sv{w})\in\Sp{W}^n$ be the configuration resulting from the player holding ball $v$ having ``given birth''.
  Thus, the $\Phi_{v}\Upsilon_{u}$ is the result of the Moran event following the random choice of an ordered pair of black balls $(u,v)$.
  Note that if either $\sv{w}'=\Phi_{u}\Upsilon_{v}(\sv{w})$ or $\sv{w}'=\Phi_{v}\Upsilon_{u}(\sv{w})$, then $\sv{w}'_{n-1}=\{u,v\}$, i.e., $u$ and $v$ are the balls held by the rightmost player immediately after the rearrangement.
  Let $\mathscr{M}(\sv{w},\sv{w}')$ be the $\sv{w}$ to $\sv{w}'$ transition probability.
  Then, supposing $\sv{w}'_{n-1}=\{a,b\}$,
  \begin{equation}
    \label{eq:M}
    \mathscr{M}(\sv{w},\sv{w}')=\sum_{\substack{u,v=1\\u\ne v}}^{n}\!\frac{\indicator{\Phi_{u}\Upsilon_{v}(\sv{w})=\sv{w}'}}{n(n-1)}=\frac{\indicator{\Phi_{a}\Upsilon_{b}(\sv{w})=\sv{w}'}+\indicator{\Phi_{b}\Upsilon_{a}(\sv{w})=\sv{w}'}}{n(n-1)},
  \end{equation}
  where $\indicator{A}$ denotes the indicator function for the condition $A$.
  One easily verifies that $\mathscr{M}$ is stochastic, i.e., $\sum_{\sv{w}'}\!\mathscr{M}(\sv{w},\sv{w}')=1$.
  In fact, $\mathscr{M}$ is doubly stochastic, i.e., $\sum_{\sv{w}}\!\mathscr{M}(\sv{w},\sv{w}')=1$.
  To see this, note that, by the reasoning of the last paragraph, for each $\sv{w}'\in\Sp{W}^n$, $\vert(\Phi_{a}\Upsilon_{b})^{-1}(\{\sv{w}'\})\vert=\binom{n}{2}$ when $\sv{w}'_{n-1}=\{a,b\}$ and $\vert(\Phi_{a}\Upsilon_{b})^{-1}(\{\sv{w}'\})\vert=0$ otherwise.
\end{proof}

\begin{prop}
  \label{prop:Mj}
  Suppose $\sv{w}'_{n-1}=\{a,b\}$.
  Let
  \begin{equation*}
    \mathscr{M}_j(\sv{w},\sv{w}')\coloneqq\frac{\indicator{\Phi_{a}\Upsilon_{b}(\sv{w})=\sv{w}'\;\&\;b\in \sv{w}_j}+\indicator{\Phi_{b}\Upsilon_{a}(\sv{w})=\sv{w}'\;\&\;a\in \sv{w}_j}}{n(n-1)}.
  \end{equation*}
  $\mathscr{M}_j(\sv{w},\sv{w}')$ is the probability that, in the move from $\sv{w}$ to $\sv{w}'$, the first black ball selected is held by the player in seat $j$.
  Clearly, $\mathscr{M}(\sv{w},\sv{w}')=\sum_{j=1}^{n-1}\!\mathscr{M}_j(\sv{w},\sv{w}')$.
  Moreover
  \begin{equation*}
    \begin{gathered}
      \sum_{\sv{w}\in\Sp{W}^n}\!\mathscr{M}_j(\sv{w},\sv{w}')=\frac{j}{\binom{n}{2}} \qquad
      \text{and} \qquad
      \sum_{\sv{w}'\in\Sp{W}^n}\!\mathscr{M}_j(\sv{w},\sv{w}')=\frac{B_{j}(\sv{w})}{n},
    \end{gathered}
  \end{equation*}
  where $B_j(\sv{w})=\vert\{a\in \sv{w}_j:a\ \text{is black}\}\vert$, i.e., the number of black balls held by the player in seat $j$.
\end{prop}
\begin{proof}
  Note that, by the counting argument above, $\Big\vert\{\sv{w}:\:\Phi_{b}\Upsilon_{a}(\sv{w})=\sv{w}'\;\&\;a\in \sv{w}_j\}\Big\vert = j$.
  Furthermore, for each $\sv{w}\in\Sp{W}^n$, the probability that the player in seat $j$ is killed is just $B_j(\sv{w})/n$.
\end{proof}

\begin{corol}
  \begin{equation*}
    \sv{W}\sim q_n \implies \expect{\frac{B_j(\sv{W})}{n}} = \frac{j}{\binom{n}{2}}
  \end{equation*}
\end{corol}


\subsection{Ergodicity}
\label{sec:uniq}

For the MGP $\sv{X}$, we define the transition probability kernel, $P^t$, in the usual fashion \citep[e.g.,][]{Feller1957,Meyn2009}.
Specifically, let
\begin{equation}
  \label{eq:Pdef}
  P^t(\sv{x},\mathcal{E})\coloneqq\prob{\sv{X}(t)\in\mathcal{E}\;\vert\;\sv{X}(0)=\sv{x}},
\end{equation}
for $t\ge 0$, $\sv{x}\in\Sp{X}^n$, and measurable $\mathcal{E}\subset\Sp{X}^n$.
For each $\sv{x}$, $P^t(\sv{x},\cdot)$ is a measure, while for each $\mathcal{E}$, $P^t(\cdot,\mathcal{E})$ is a measurable function from $\Sp{X}^n$ to $[0,1]$.
Moreover, for every $\sv{x}$ and $\mathcal{E}$, $t\mapsto P^t(\sv{x},\mathcal{E})$ is a measurable function.

Define the family of operators, $K^t$, $t\ge 0$, by
\begin{equation}
  \label{eq:Kdef}
  (K^t f)(\sv{x})\coloneqq\expect{f(\sv{X}(t))\;\vert\;\sv{X}(0)=\sv{x}}=\int_{\Sp{X}^n}\!P^t(\sv{x},\dd{\sv{x}'})\,f(\sv{x}'),
\end{equation}
for $\sv{x}\in\Sp{X}^n$ and $f\in\mathcal{D}\coloneqq\{g:\Sp{W}^n\times\Sp{S}^n\to\Sp{R}\;\vert\;\forall\sv{w}\ g(\sv{w},\cdot)\in C^1(\Sp{S}^n)\}$.
With this definition, $K^t$ is a Markov semigroup.
Note that the adjoint action of $K^t$ is naturally defined, as follows.
Let $\nu$ be a measure on $\Sp{X}^n$.
Then, for all $t\ge 0$, the measure $\nu K_t$ is defined by
\begin{equation}
  \label{eq:Kadj}
  (\nu K^t)(\mathcal{E})\coloneqq\int_{\Sp{X}^n}\!\nu(\dd{\sv{x}})\,P^t(\sv{x},\mathcal{E}),
\end{equation}
for all measurable $\mathcal{E}\subset\Sp{X}^n$.

\begin{thm}
  \label{thm:ergodic}
  The Moran genealogy process is uniformly and exponentially ergodic.
  In particular, there are constants $D<\infty$ and $0\le\rho<1$ such that
  \begin{equation*}
    \left\Vert \nu K^t-\pi_n\right\Vert_{\mathrm{TV}}<D\,\rho^t,
  \end{equation*}
  for all $t\ge 0$ and all probability measures $\nu$ on $\Sp{X}^n$.
  Here, the norm on measures is the total variation norm, defined by
  \begin{equation*}
    \left\Vert\mu\right\Vert_{\mathrm{TV}}\coloneqq\sup_{|f|\le 1}\left\vert\mu(f)\right\vert.
  \end{equation*}
  Moreover, if $\opnorm{\cdot}_{\infty}$ denotes the $L^{\infty}$ operator norm, with the same $D$ and $\rho$ as above, we have
  \begin{equation*}
    \opnorm{K^t-\pi_n}_{\infty}<D\,\rho^t,\quad\text{for}\ t\ge 0.
  \end{equation*}
\end{thm}
\begin{proof}
  We establish ergodicity by studying the resolvent, $U_1$, of $K^t$:
  \begin{equation}
    \label{eq:resolvent}
    U_1\coloneqq\int_0^\infty\!e^{-t}\,K^t\,\dd{t}.
  \end{equation}
  $U_1$ is itself the generator of the Markov semigroup of the discrete-time Markov chain obtained by observing $\sv{X}$ at a random sequence of times.
  Specifically, let $R_k$ be a unit-rate Poisson process on $\mathbb{R}_{+}$, independent of $\sv{X}$.
  Define the \emph{resolvent chain}, $\sv{Y}_k\coloneqq\sv{X}(R_k)$, $k\in\mathbb{N}$.
  Then $U_1^k$ is the Markov semigroup of the chain $\{\sv{Y}_k\}_{k\in\mathbb{N}}$.

  Now let $\dd{\sv{s}}=\prod_{j=1}^{n-1}\dd{\sv{s}_j}$ denote Lebesgue measure on $\Sp{S}^n$ and $\dd{\sv{w}}$ be counting measure on $\Sp{W}^n$.
  Define the finite measure $\eta$ on $\Sp{X}^n=\Sp{W}^n\times\Sp{S}^n$ by
  \begin{equation}
    \label{eq:etadef}
    \eta(\dd{\sv{w}}\,\dd{\sv{s}}) \coloneqq \frac{1}{2^{n-1}}\,\exp\left[-(1+\mu)\,\left(\sv{s}_1+\cdots+\sv{s}_{n-1}\right)\right]\,\dd{\sv{w}}\,\dd{\sv{s}},
  \end{equation}
  where we recall that $\mu$ is the Moran event-rate.
  To be clear, \cref{eq:etadef} defines $\eta$ as a product measure, where the $\Sp{W}^n$-component is counting measure and the $\Sp{S}^n$-component is absolutely continuous with respect to Lebesgue measure.

  Next, suppose $\sv{x}\in\Sp{X}^n$ is an arbitrary state and $\mathcal{E}$ is a measurable subset of $\Sp{X}^n$.
  Note that for any $(\sv{w},\sv{s})\in\mathcal{E}$, there is a sample path that leads directly from $\sv{x}$ to $(\sv{w},\sv{s})$ in precisely $\sv{s}_1+\dots+\sv{s}_{n-1}$ units of time and with exactly $n-1$ events transpiring at intervals $\sv{s}_1,\dots,\sv{s}_{n-1}$.
  At each event, there is at least one choice of a pair of black balls that can be made.
  Hence the probability associated with any of these paths is
  \begin{equation*}
    \frac{1}{2^{n-1}}\,\exp\left[-\mu\,\sum_{m=1}^{n-1}\!\sv{s}_m\right]\dd{\sv{s}}.
  \end{equation*}
  The probability that $R_{k+1}-R_k=\sum\!\sv{s}_m$ is $\exp(-\sum\!\sv{s}_m)\,\dd{\sv{s}}$.
  Summing over all $(\sv{w},\sv{s})$ in $\mathcal{E}$ gives
  \begin{equation}
    \label{eq:harris}
    \prob{\sv{Y}_k\in\mathcal{E}\;\vert\;\sv{Y}_{k-1}=\sv{x}} \ge \eta(\mathcal{E})
  \end{equation}
  independent of $\sv{x}$ and $k$.
  Though we do not use it, in fact the inequality is strict since, although the constructed paths are almost surely those which reach $(\sv{w},\sv{s})$ in minimal time, there are many others that arrive by more circuitous routes.
  Since \cref{eq:harris} holds for all $\sv{x}\in\Sp{X}^n$, the full state space $\Sp{X}^n$ is said to be a \emph{petite set}.
  It follows from Theorem~16.2.2 of \citet{Meyn2009} that the resolvent chain $\sv{Y}$ is uniformly ergodic and therefore, \emph{a fortiori}, that $\sv{Y}$ possesses a unique invariant distribution, $\pi_n$, i.e., one for which $\pi_n=\pi_n U_1$.
  Since $\eta$ is finite, moreover, $\pi_n$ can be taken to be a probability distribution.
  By a lemma of \citet{Azema1967} \citep[which they attribute to][]{Nagasawa1963}, a measure is invariant under $U_1$ if and only if it is invariant under $K^t$, for all $t$.
  It follows that $\pi_n$ is the unique invariant probability measure of $\sv{X}$ as well.


  To establish the ergodicity of $\sv{X}$, we verify that the $\sv{Y}$ satisfies a drift condition.
  In particular, \citet[][Theorem~5.2]{Down1995} show that
  if, for some petite set $\mathcal{C}\subset\Sp{X}^n$, some function $V:\Sp{X}^n\to [1,\infty)$, some $\lambda<1$, and some $b<\infty$,
    one has $\expect{V(\sv{Y_k})|\sv{Y}_0=\sv{x}}\le\lambda\,V(\sv{x})+b\,\indicator{\sv{x}\in\mathcal{C}}$ for every $\sv{x}$,
    then one can conclude that $\sv{X}$ is $V$-uniformly ergodic.
    We take $V(\sv{x})=1$, $\lambda=b=\frac{1}{2}$, and $\mathcal{C}=\Sp{X}^n$.
    We conclude that $\sv{X}$ is uniformly ergodic in the sense of \citet{Down1995}:
    this is equivalent to the first statement in the theorem.

    Finally, it is easy to see that
    \begin{equation*}
      \left\Vert K^tf-\pi_n(f)\right\Vert_{\infty}<D\,\left\Vert f \right\Vert_{\infty}\,\rho^t, \quad t\ge 0, \quad f\in\mathcal{D}.
    \end{equation*}
    The second statement in the theorem then follows immediately from the definition of the operator norm.
\end{proof}

We determine the form of $\pi_n$ in \cref{sec:inv_meas}.

\subsection{Infinitesimal generator}

With the Markov semigroup $K^t$ defined by \cref{eq:Kdef}, we have
\begin{equation*}
  (K^t f)(\sv{w},\sv{s})=\int_{\Sp{W}^n\times\Sp{S}^n}\!\mathscr{P}(t,\sv{w},\sv{s},\sv{w}',\sv{s}')\,f(\sv{w}',\sv{s}')\,\dd{\sv{w}'}\,\dd{\sv{s}'}+o(t),
\end{equation*}
as $t\downarrow 0$, where
\begin{gather}
  \mathscr{P}(t,\sv{w},\sv{s},\sv{w}',\sv{s}')\coloneqq e^{-\mu t}\,\mathscr{Q}(t,\sv{w},\sv{s},\sv{w}',\sv{s}') + \left(1-e^{-\mu t}\right)\,\sum_{j=1}^{n-1}\mathscr{M}_j(\sv{w},\sv{w}')\,\mathscr{R}_j(\sv{s},\sv{s'}),\label{eq:K}\\
  \mathscr{Q}(t,\sv{w},\sv{s},\sv{w}',\sv{s}')\coloneqq\delta(\sv{s}_{n-1}+t-\sv{s}'_{n-1})\cdot\prod_{j=1}^{n-2}\!\delta(\sv{s}_{j}-\sv{s}'_{j})\cdot\indicator{\sv{w}=\sv{w}'},\,\text{and}\\
  \mathscr{R}_j(\sv{s},\sv{s}')\coloneqq\prod_{k=1}^{j-2}\!\delta(\sv{s}_k-\sv{s}'_k)\cdot\delta(\sv{s}_{j-1}+\sv{s}_{j}-\sv{s}'_{j-1})\cdot\prod_{k=j}^{n-2}\!\delta(\sv{s}_{k+1}-\sv{s}'_{k})\cdot\delta(\sv{s}'_{n-1}).\label{eq:Rj}
\end{gather}
Here, $\delta$ is the Dirac delta function.
The $\mathscr{Q}$ term encodes changes in $\sv{X}$ that occur between birth-death events:
the $(n-1)$-st coalescent interval grows with time, while the other intervals remain fixed and the topology remains unchanged.
The $\mathscr{M}_j\mathscr{R}_j$ terms in the sum of \cref{eq:K} encode the changes in $\sv{X}$ that occur when the selected black ball is held by the player in seat $j$.
At such an event, $\sv{w}$ jumps to $\sv{w}'$ with probability $\mathscr{M}_j(\sv{w},\sv{w}')$, the $(j-1)$-st coalescent interval subsumes the $j$-th, while the $k$-th interval takes the value of the $(k-1)$-st for $k\ge j$.
Moreover, the $(n-1)$-st interval is set to zero.

We compute the infinitesimal generator, $L$, as the linear operator satisfying
\begin{equation*}
  \lim_{t\;\downarrow\;0}\frac{K^t f - f}{t} = L f
\end{equation*}
for $f\in\mathcal{D}$.
This is easily done, and we obtain the following, which we state without proof.
\begin{prop}\label{prop:generator}
  The infinitesimal generator of the MGP is the linear operator $L$ defined by
  \begin{equation}\label{eq:generator1}
    (L f)(\sv{w},\sv{s}) = \int_{\Sp{W}^n\times\Sp{S}^n}\!\mathscr{L}(\sv{w},\sv{s},\sv{w}',\sv{s}')\,f(\sv{w}',\sv{s}')\,\dd{\sv{w}'}\,\dd{\sv{s}'},
  \end{equation}
  whenever $f\in\mathcal{D}$.
  The kernel, $\mathscr{L}$, is given by
  \begin{equation}
    \label{eq:generator}
    \begin{split}
      \mathscr{L}(\sv{w},\sv{s},\sv{w}',\sv{s}') \coloneqq &\left(\delta'(\sv{s}_{n-1}-\sv{s}'_{n-1})-{\mu}\,\delta(\sv{s}_{n-1}-\sv{s}'_{n-1})\right)\,\prod_{k=1}^{n-2}\!\delta(\sv{s}_k-\sv{s}'_k)\,\indicator{\sv{w}=\sv{w}'}\\&+\mu\,\sum_{j=1}^{n-1}\mathscr{M}_j(\sv{w},\sv{w}')\,\mathscr{R}_{j}(\sv{s},\sv{s}').
    \end{split}
  \end{equation}
  Here, the symbol $\delta'$ refers to the derivative of the Dirac delta function.
\end{prop}

\subsection{Kolmogorov backward equation}

For $f\in\mathcal{D}$, let $u(t,\sv{w},\sv{s})\coloneqq K^tf(\sv{w},\sv{s})$.
Note that
\begin{equation}\label{eq:KBE0}
  \begin{split}
    \pd{u}{t}(t,\sv{w},\sv{s})&=\lim_{\Delta{t}\downarrow 0}\frac{K^{t+\Delta{t}}f(\sv{\sv{w},\sv{s}})-K^tf(\sv{w},\sv{s})}{\Delta{t}}=\lim_{\Delta{t}\downarrow 0}\frac{K^{\Delta{t}}-\mathrm{Id}}{\Delta{t}}\,K^tf(\sv{\sv{w},\sv{s}})\\
    &=L\,u(t,\sv{w},\sv{s})=\int_{\Sp{W}^n\times\Sp{S}^n}\mathscr{L}(\sv{w},\sv{s},\sv{w}',\sv{s}')\,u(t,\sv{w}',\sv{s}')\,\dd{\sv{w}'}\,\dd{\sv{s}'}.
  \end{split}
\end{equation}
Let the functions $\mathbf{\sigma}_j:\Sp{S}^n\to\Sp{S}^n$ be defined by
\begin{equation*}
  \begin{aligned}
    &\mathbf{\sigma}_1(\sv{s}_1,\dots,\sv{s}_{n-1})\coloneqq\left(\sv{s}_2,\dots,\sv{s}_{n-1},0\right)\\
    &\mathbf{\sigma}_j(\sv{s}_1,\dots,\sv{s}_{n-1})\coloneqq\left(\sv{s}_1,\dots,\sv{s}_{j-2},\sv{s}_{j-1}+\sv{s}_j,\sv{s}_{j+1},\dots,\sv{s}_{n-1},0\right), \qquad j=2,\dots,n-1.
  \end{aligned}
\end{equation*}
Applying \cref{eq:generator} to \cref{eq:KBE0}, we obtain the Kolmogorov backward equation,
\begin{equation}
  \label{eq:KBE1}
  \frac{1}{\mu}\,\pd{u}{t}(t,\sv{w},\sv{s})-\frac{1}{\mu}\,\pd{u}{\sv{s}_{n-1}}(t,\sv{w},\sv{s})=
  \sum_{j=1}^{n-1}\sum_{\sv{w}'\in\Sp{W}^n}\!\mathscr{M}_j(\sv{w},\sv{w}')\,\left[u(t,\sv{w}',\mathbf{\sigma}_j(\sv{s}))-u(t,\sv{w},\sv{s})\right].
\end{equation}
Together with the initial condition, $u(0,\sv{w},\sv{s})=f(\sv{w},\sv{s})$, \cref{eq:KBE1} determines the Markov semigroup $K^t$.
In the special case that $f$ is independent of $\sv{w}$, we can average \cref{eq:KBE1} over $\sv{w}$ and apply \cref{prop:Mj} to obtain,
\begin{equation}
  \label{eq:KBE2}
  \frac{1}{\mu}\,\pd{u}{t}(t,\sv{s})-\frac{1}{\mu}\,\pd{u}{\sv{s}_{n-1}}(t,\sv{s})=
  \sum_{j=1}^{n-1}\frac{j}{\binom{n}{2}}\,\left[u(t,\mathbf{\sigma}_j(\sv{s}))-u(t,\sv{s})\right].
\end{equation}


\subsection{Invariant measure}\label{sec:inv_meas}

In this section, it will be convenient to scale time so that $\mu={\binom{n}{2}}$.

The invariant measure, $\pi_n$, of the MGP is characterized by the fact that it is annihilated by the generator, i.e., $\pi_n L=0$.
We seek a separable measure $\pi_n(\dd{\sv{w}}\,\dd{\sv{s}})=p_0(\sv{w})\,\prod_{k=1}^{n-1} p_k(\sv{s}_k)\,\dd{\sv{w}}\,\dd{\sv{s}}$.
Operating with $L$ on $\pi_n$ involves integrating over all possible genealogies $(\sv{w},\sv{s})$:
\begin{equation}
  \label{eq:Lpi}
  \begin{split}
    \pi_n L = &\int_{\Sp{W}^n\times\Sp{S}^n}\!\pi_n(\dd{\sv{w}}\,\dd{\sv{s}})\,\mathscr{L}(\sv{w},\sv{s},\sv{w}',\sv{s}')\\
    =-&\left(p'_{n-1}(\sv{s}'_{n-1})+\binom{n}{2}\,p_{n-1}(\sv{s}'_{n-1})\right)\,\prod_{k=1}^{n-2}\!p_k(\sv{s}'_k)\,p_0(\sv{w}')\\
    &+\binom{n}{2}\,\sum_{j=1}^{n-1}\!\sum_{\sv{w}}\!p_0(\sv{w})\,\mathscr{M}_j(\sv{w},\sv{w}')\,\prod_{k=1}^{j-2}\!p_k(\sv{s}'_k)\cdot Q_{j-1}(\sv{s}'_{j-1})\cdot\prod_{k=j}^{n-2}\!p_{k+1}(\sv{s}'_{k})\cdot\delta(\sv{s}'_{n-1}) = 0.
  \end{split}
\end{equation}
Here $p'_{n-1}\coloneqq\partial{p_{n-1}}/\partial{\sv{s}_{n-1}}$ and $Q_j$ is defined by
\begin{equation}
  \label{eq:Qdef}
  Q_j(s) \coloneqq \int_0^{s}\,p_j(t)\,p_{j+1}(s-t)\,\dd{t}.
\end{equation}

Integrating out all the $\sv{s}_j$ in \cref{eq:Lpi}, we obtain the matrix equation
\begin{equation*}
  p_0(\sv{w}')=\sum_{\sv{w}}\!p_0(\sv{w})\,\sum_{j=1}^{n-1}\!\mathscr{M}_j(\sv{w},\sv{w}')=\sum_{\sv{w}}\!p_0(\sv{w})\,\mathscr{M}(\sv{w},\sv{w}'),
\end{equation*}
which is just the expression of the requirement that $p_0$ be the stationary distribution of the $\sv{W}$ chain, which we have already determined:
indeed, \cref{prop:q} states that $p_0(\sv{w})=\mathrm{const}$.

To find the other factors of $\pi_n$, we divide both sides of \cref{eq:Lpi} by $p_0(\sv{w})=p_0(\sv{w}')$ and, after dropping the primes, which are no longer needed, we have
\begin{equation}
  \label{eq:Lpi2}
  \begin{split}
    -&\left(p'_{n-1}(\sv{s}_{n-1})+\binom{n}{2}\,p_{n-1}(\sv{s}_{n-1})\right)\,\prod_{k=1}^{n-2}\!p_k(\sv{s}_k)\\
    &\qquad\qquad+\sum_{j=1}^{n-1}\,j\,\prod_{k=1}^{j-2}\!p(\sv{s}_k)\cdot Q_{j-1}(\sv{s}_{j-1})\cdot\prod_{k=j}^{n-2}\!p_{k+1}(\sv{s}_{k})\cdot\delta(\sv{s}_{n-1}) = 0.
  \end{split}
\end{equation}
Note that, in passing from \cref{eq:Lpi} to \cref{eq:Lpi2}, we have applied \cref{prop:Mj}.

Since \Cref{eq:Lpi2} holds for all $\sv{s}_{n-1}$, we must have,
\begin{equation*}
  p'_{n-1}(\sv{s}_{n-1})+\binom{n}{2}\,p_{n-1}(\sv{s}_{n-1})=0,
\end{equation*}
for $\sv{s}_{n-1}>0$, whence
\begin{equation}
  \label{eq:case_n}
  p_{n-1}(s) = \binom{n}{2}\,e^{-\binom{n}{2}\,s}.
\end{equation}

Now, we integrate \cref{eq:Lpi} over $\sv{s}_{n-1}$, which yields
\begin{equation}
  \label{eq:topface}
  -\binom{n}{2}\,\prod_{k=1}^{n-2}\!p_k(\sv{s}_k)+\sum_{j=1}^{n-1}\,j\,\prod_{k=1}^{j-2}\!p_k(\sv{s}_k)\cdot Q_{j-1}(\sv{s}_{j-1})\cdot\prod_{k=j}^{n-2}\!p_{k+1}(\sv{s}_{k})=0.
\end{equation}
Notice that each term in the sum of \cref{eq:topface} contains a product of $n-2$ factors, each of which is a probability density over a different one of the $\sv{s}_1,\dots,\sv{s}_{n-2}$ variables.
Consequently, by integrating over all $\sv{s}_k$, $k\ne m$, we obtain an expression for the marginal density of $\sv{s}_m$:
\begin{equation}
  \label{eq:Smarg}
  p_{m+1}(\sv{s}_m)+\frac{2}{m}\,Q_m(\sv{s}_m)-\left(1+\frac{2}{m}\right)p_{m}(\sv{s}_m)=0,
\end{equation}
which holds for $m=1,\dots,n-2$.

We establish, by reverse induction on $m$, that $p_m(s)=\binom{m+1}{2}\,e^{-\binom{m+1}{2}\,s}$ for $m=1,\dots,n-1$.
In \cref{eq:case_n}, we have already shown the result for $m=n-1$.
Applying the ${{\partial}/{\partial \sv{s}_m}}+\binom{m+2}{2}$ operator to both sides of \cref{eq:Smarg} yields
\begin{equation}
  \label{eq:Smarg2}
  \begin{split}
    &\left(p'_{m+1}+\binom{m+2}{2}\,p_{m+1}\right)+\frac{2}{m}\,\left(Q'_m+\binom{m+2}{2}\,Q_m\right)\\
    &\qquad\qquad\qquad\qquad\qquad-\left(1+\frac{2}{m}\right)\,\left(p'_{m}+\binom{m+2}{2}\,p_{m}\right)=0.
  \end{split}
\end{equation}
By the induction hypothesis, the first term of \cref{eq:Smarg2} vanishes and the second term simplifies, and we are left with
\begin{equation*}
  \label{eq:Smarg3}
  p'_{m}+\frac{m}{m+2}\,\binom{m+2}{2}\,p_m=0.
\end{equation*}
The result follows.

We have now established that, when $\mu=\binom{n}{2}$,
\begin{equation*}
  \pi_n(\dd{\sv{w}}\,\dd{\sv{s}})=\exp\left(-\sum_{j=1}^{n-1}\!\binom{j+1}{2}\,\sv{s}_j\right)\,\dd{\sv{w}}\,\dd{\sv{s}}.
\end{equation*}
More generally, we have
\begin{thm}
  \label{thm:pi}
  If the MGP of size $n$ proceeds with event rate $\mu$, then its unique invariant probability measure, $\pi_n$, is given by
  \begin{equation*}
    \pi_n(\dd{\sv{w}}\,\dd{\sv{s}}) = \left({{\mu}\over{\smash{\binom{n}{2}}}}\right)^{n-1}\,\exp\left(-\sum_{j=1}^{n-1}\!\frac{\binom{j+1}{2}}{\binom{n}{2}}\,\mu\,\sv{s}_j\right)\,\dd{\sv{w}}\,\dd{\sv{s}},
  \end{equation*}
  where $\dd{\sv{w}}$ is the counting measure on $\Sp{W}^n$ and $\dd{\sv{s}}$ is Lebesgue measure on $\Sp{S}^n$.
\end{thm}

Thus, the unique limiting stationary measure of the Moran Genealogy Process is identical to the probability measure of the \citet{Kingman1982} coalescent.
Although this result is unsurprising, the proof, which is constructive and strictly forward-looking, sheds additional light onto the relationship between the Moran process and the Kingman coalescent.

\subsection{Synchronous sampling}\label{sec:sampling}

We now ask about the probability of sampling a given genealogy.
Specifically, we imagine that at a given time, we sample $k$ individuals from the population at random.
What is the genealogy linking these individuals?

We can represent the process of sampling a subgenealogy of size $k$ in terms of the Moran genealogy game as follows.
Specifically, we perform the following iterative procedure.
\begin{inparaenum}[(1)]
\item The player holding the highest-numbered black ball stands up.
\item He exchanges his second ball for the green ball bearing his name.
\item The standing player is dismissed, all players seated to his right shift one seat to the left, and the rightmost chair is removed.
\end{inparaenum}
At each iteration of this procedure, we remove the highest-numbered black ball from play and dismiss one player.
Thus after $n-k$ steps, the configuration of the balls held by the remaining $k$ players, and the numbers written on their slates, together determine the sampled genealogy.
Each step in this procedure kills one player, sequentially applying the function $\Upsilon_u$ (defined in \cref{sec:w-dist}) for $u=n,n-1,\dots,k+1$.
Since $\Upsilon_u\,\Upsilon_v=\Upsilon_v\,\Upsilon_u$ for all $u$, $v$, the result would be the same were we to kill players by announcing a random sequence of $n-k$ black balls, provided we then replaced the remaining black balls with those numbered $1,\dots,k$.

For $\sv{x}\in\Sp{X}^n$, let $\Omega_n(\sv{x})\in\Sp{X}^{n-1}$ represent the random result of drawing a sample of size $n-1$ from $\sv{x}\in\Sp{X}^{n-1}$ as just described.
The selection of a sample of size $k$ is then the $(n-k)$-fold composition, $\Omega^k_n\coloneqq\Omega_{k+1}\circ\Omega_{k+2}\circ\cdots\circ\Omega_{n}$.
We risk no confusion in defining $\Omega_n(\sv{w})\in\Sp{W}^{n-1}$ to be the corresponding projection of $\Omega_n(\sv{x})$ onto its $\Sp{W}^{n-1}$-component.

The removal of each successive player (i.e., application of the random function $\Omega_m$) is itself equivalent to an application of the deterministic function $\Upsilon_{u}$ (defined in \cref{sec:w-dist}) for some randomly selected black ball $u$.
For $\sv{w}\in\Sp{W}^n$ and $\sv{w}'\in\Sp{W}^{n-1}$, let $\mathscr{H}^n(\sv{w},\sv{w}')$ be the probability that $\Omega_n(\sv{w})=\sv{w}'$ and
denote by $a$ the unique black ball in $\cup_{j=1}^{n-1} \sv{w}_j\setminus\cup_{j=1}^{n-2} \sv{w}'_j$.
Then
\begin{equation*}
  \mathscr{H}^n(\sv{w},\sv{w}')=\frac{1}{n}\sum_{u=1}^n\!\indicator{\Upsilon_u(\sv{w})=\sv{w}'}=\frac{\indicator{\Upsilon_a(\sv{w})=\sv{w}'}}{n}.
\end{equation*}
A counting argument similar to that employed in \cref{sec:w-dist} shows that
\begin{equation}
  \sum_{\sv{w}\in\Sp{W}^n}\mathscr{H}^n(\sv{w},\sv{w}') = \binom{n}{2},\ \text{for all}\ \sv{w}'\in\Sp{W}^{n-1}.
\end{equation}
As in \cref{prop:Mj}, we decompose $\mathscr{H}^n$ into $n-1$ levels, writing
$\mathscr{H}^n(\sv{w},\sv{w}')=\sum_{j=1}^{n-1}\!\mathscr{H}^n_j(\sv{w},\sv{w}')$, where
\begin{equation}
  \mathscr{H}^n_j(\sv{w},\sv{w}')\coloneqq\frac{\indicator{\Upsilon_a(\sv{w})=\sv{w}'\;\&\;a\in \sv{w}_j}}{n}.
\end{equation}
Again, simple counting arguments establish that
\begin{equation}
  \begin{gathered}
    \sum_{\sv{w}\in\Sp{W}^n}\mathscr{H}^n_j(\sv{w},\sv{w}')=j\qquad
    \text{and} \qquad
    \sum_{\sv{w}'\in\Sp{W}^{n-1}}\!\mathscr{H}^n_j(\sv{w},\sv{w}')=\frac{B_{j}(\sv{w})}{n},
  \end{gathered}
\end{equation}
where $B_j$ is as defined in \cref{prop:Mj}.

What is the action of the sampling operation $\Omega_n$ on probability measures?
Given any probability measure $\nu$ on $\Sp{X}^n$ and any event $\mathcal{E}\subset\Sp{X}^{n-1}$, define
\begin{equation}
  \label{eq:psinu}
  \nu\Omega_n(\mathcal{E}) \coloneqq \int_{\mathcal{E}}\!\int_{\Sp{X}^n}\!\nu(\dd{\sv{w}}\,\dd{\sv{s}})\,\sum_{j=1}^{n-1}\mathscr{H}^n_j(\sv{w},\sv{w}')\,\mathscr{R}^n_j(\sv{s},\sv{s}')\,\dd{\sv{w}'}\,\dd{\sv{s}'}.
\end{equation}
Here, $\mathscr{R}^n_j$ is defined in a manner similar to \cref{eq:Rj}, by
\begin{equation*}
  \mathscr{R}^n_j(\sv{s},\sv{s}')\coloneqq\prod_{k=1}^{j-2}\!\delta(\sv{s}_k-\sv{s}'_k)\cdot\delta(\sv{s}_{j-1}+\sv{s}_{j}-\sv{s}'_{j-1})\cdot\prod_{k=j}^{n-2}\!\delta(\sv{s}_{k+1}-\sv{s}'_{k}),
\end{equation*}
for $\sv{s}\in\Sp{S}^n$, $\sv{s}'\in\Sp{S}^{n-1}$.
Again, without loss of generality, we scale time so that $\mu=\binom{n}{2}$.
Applying \cref{eq:psinu} to the stationary measure, $\pi_n$, of the size-$n$ MGP (\cref{thm:pi}), we obtain
\begin{equation*}
  \pi_n\Omega_n(\dd{\sv{w}'}\dd{\sv{s'}})=\mathscr{I}(\sv{w}',\sv{s}')\,\dd{\sv{w}'}\dd{\sv{s'}},
\end{equation*}
where the density $\mathscr{I}$ satisfies
\begin{equation}
  \label{eq:psipi}
  \begin{split}
    \mathscr{I}(\sv{w}',\sv{s}')=&\sum_{j=1}^{n-1}\int_{\Sp{W}^{n}\times\Sp{S}^{n}}\!q_n(\sv{w})\,\mathscr{H}^n_j(\sv{w},\sv{w}')\,\mathscr{R}^n_j(\sv{s},\sv{s}')\,\prod_{m=1}^{n-1}\!p_m(\sv{s}_m)\,\dd{\sv{w}}\,\dd{\sv{s}}\\
    =&\sum_{j=1}^{n-1}\left(\sum_{\sv{w}\in\Sp{W}^n}\!q_n(\sv{w})\,\mathscr{H}^n_j(\sv{w},\sv{w}')\right)\,\left(\int_{\Sp{S}^{n}}\!\mathscr{R}^n_j(\sv{s},\sv{s}')\,\prod_{m=1}^{n-1}\!p_m(\sv{s}_m)\,\dd{\sv{s}}\right)\\
    =&\sum_{j=1}^{n-1}\frac{j\,2^{n-1}}{n!(n-1)!}\,\prod_{m=1}^{j-2}\!p_m(\sv{s}'_m)\cdot Q_{j-1}(\sv{s}'_{j-1})\cdot\prod_{m=j}^{n-2}\!p_{m+1}(\sv{s}'_m).
  \end{split}
\end{equation}
Here, as before,  $q_n(\sv{w})\coloneqq\tfrac{2^{n-1}}{n!\,(n-1)!}$, $p_m(s)\coloneqq\binom{m+1}{2}\,\exp\left(-\binom{m+1}{2}s\right)$ and, from \cref{eq:Qdef},
\begin{equation*}
  Q_{j-1}(s) = \binom{j}{2}\,\binom{j+1}{2}\,\frac{\exp\left(-\binom{j}{2}s\right)-\exp\left(-\binom{j+1}{2}s\right)}{j}.
\end{equation*}
Substituting these expressions into \cref{eq:psipi} and doing some routine algebra gives
\begin{equation*}
  \mathscr{I}(\sv{w},\sv{s})=\exp\left(-\sum_{j=1}^{n-2}\!\binom{j+1}{2}\sv{s}_j\right),
\end{equation*}
which implies that $\pi_n\Omega_n=\pi_{n-1}$.
Iterating this result $n-k$ times establishes
\begin{thm}\label{thm:xsample}
  Let $\sv{X}_n$ be the stationary Moran genealogy process of size $n$ and event rate $\mu$.
  For $k\le n$, let $\sv{Z}_k=\Omega^k_n(\sv{X}_n)$ be the corresponding size-$k$ sampled process.
  Then the marginal probability distribution of $\sv{Z}_k(t)$ on $\Sp{X}^k$ is given by the measure
  \begin{equation*}
    \left({{\mu}\over{\smash{\binom{n}{2}}}}\right)^{k-1}\,\exp\left(-\sum_{j=1}^{k-1}\!\frac{\binom{j+1}{2}}{\binom{n}{2}}\,\mu\,\sv{s}_j\right)\,\dd{\sv{w}}\,\dd{\sv{s}},
  \end{equation*}
  where, as before, $\dd{\sv{w}}$ is the counting measure on $\Sp{W}^n$ and $\dd{\sv{s}}$ is Lebesgue measure on $\Sp{S}^n$.
\end{thm}

\end{document}